\documentclass[10pt,journal]{IEEEtran}

\usepackage{amsmath,amssymb,amsfonts,amsthm,color}
\usepackage{cite}

\usepackage[pdftex]{graphicx}
\usepackage[hyphens]{url}
\usepackage[caption=false,font=footnotesize]{subfig}
\usepackage{stfloats}
\usepackage{tikz}
\usetikzlibrary{positioning, shapes, decorations.pathreplacing, decorations.markings, arrows}

\newcommand{\EX}{\ensuremath{\mathbb{E}}}
\newcommand{\PR}{\ensuremath{\mathbb{P}}}
\newcommand{\N}{\ensuremath{\mathbb{N}}}
\newcommand{\R}{\ensuremath{\mathbb{R}}}

\newtheorem{theorem}{Theorem}
\newtheorem{corollary}{Corollary}

\newtheorem{claim}{Claim}

\begin{document}
\title{Expected Extinction Times of Epidemics with State-Dependent Infectiousness}
\author{Akhil~Bhimaraju, 
Avhishek~Chatterjee, 
and~Lav~R.~Varshney,~\IEEEmembership{Senior~Member,~IEEE}%
\thanks{A.~Bhimaraju and L.~R.~Varshney are with the 
Coordinated Science Laboratory and the 
Department of Electrical and Computer Engineering, 
University of Illinois Urbana-Champaign, Urbana, IL, USA
(e-mail: akhilb3@illinois.edu and varshney@illinois.edu).
A.~Chatterjee is with the Department of Electrical Engineering, 
Indian Institute of Technology Madras, Chennai, India
(e-mail: avhishek@ee.iitm.ac.in).}%
\thanks{This work was supported in part by 
NSF grant ECCS-2033900, and the Center for Pathogen Diagnostics 
through the ZJU-UIUC Dynamic Engineering Science Interdisciplinary 
Research Enterprise (DESIRE).}}

\maketitle

\begin{abstract}
We model an epidemic where the per-person infectiousness in a
network of geographic localities
changes with the total number of active cases.
This would happen as people adopt more stringent 
non-pharmaceutical precautions when the
population has a larger number of active cases.
We show that there exists a sharp threshold such that when the curing rate
for the infection
is above this threshold, the expected time for the epidemic to die out is
logarithmic in the initial infection size, whereas when the curing rate is
below this threshold, the expected time for epidemic extinction is infinite.
We also show that when the per-person infectiousness goes to 
zero asymptotically as a function of the number of active cases, the expected
extinction times all have the same asymptote independent of network structure.
We make no mean-field assumption while deriving these results.
Simulations on real-world network topologies
bear out these results, while also demonstrating that if the
per-person infectiousness is large when the epidemic size is small
(i.e., the precautions are lax when the epidemic is small and only get
stringent after the epidemic has become large),
it might take a very long time for the epidemic to die out.
We also provide some analytical insight into these observations.
\end{abstract}

\begin{IEEEkeywords}
Epidemic  modeling, network analysis
\end{IEEEkeywords}

\section{Introduction}
\label{sec:intro}

Newly emerging infectious diseases that quickly spread
across population centers in an increasingly interconnected world
form a large portion of human infections \cite{MorensF2013}.
These epidemics spread over contact networks and the characteristics
of this spread have been widely studied
\cite{ShirleyR2005, Newman2002, 
GiordanoBBCDDC2020, ColizzaBBV2006, KuchlerRS2020,
ZhouWZSYHZOPS2020, GomezABMM2010}.
In this work, we develop a state-dependent infectiousness
model for the spread of epidemics 
over a network of population centers and
analytically prove that the epidemic dynamics follow certain properties.
Specifically, we characterize the expected time of epidemic extinction and
show that it exhibits a threshold behavior where it
is either logarithmic in the initial infection size or infinite depending
on whether the curing rate is higher or lower than a threshold.
We make no mean-field assumption while deriving this threshold.
We believe our model captures important features of epidemic spreading
not captured in prior literature, and our results advance the understanding
of epidemic spread.

We model  the epidemic as a Markov spreading process over a network whose
nodes represent population centers such as cities or large communities,
and the connections between them indicate the amount of contact between the
population centers.
New infections could either be due to interactions with people from
neighboring population centers, or due to community spread within the
population center.
We model these two components of epidemic spread separately.

In a typical  epidemic, especially in the early stages of newly emerging
infections, vaccines and other  pharmaceutical means to combat the disease
are unlikely to be available.
Further, in the early stages of the epidemic, the number of susceptible
people in a typical population center is very large, and effectively
infinite, until a large majority of the population has developed 
herd immunity.
We capture these properties in a model
where the number of infections in each population center can potentially
grow without bound.

{In cases where the infected population is a significant
fraction of the total population, the epidemic would spread more slowly
than what is predicted by our model.
This is because for a given number of infected individuals,
our model assumes that the susceptible population is larger than
it actually is.
So our model would over-estimate the effective rate at which the
contagion spreads, and the number of (new) infections in our model
stochastically dominates the actual number of infections.}
Thus, in those settings, the threshold obtained from our model would still
hold for the quick-extinction case.

Whereas models at the person level \cite{GaneshMT2005, FagnaniZ2017,
VanMieghemOK2008, SahnehVMS2017, SahnehCS2012} capture interactions 
between individual
people and might help us predict the probability of
a particular person getting infected, it is prohibitively expensive to
collect information about all individuals in a city
and compute over a network that treats each person as a distinct node.
Population~center-level models allow us to predict the
epidemic trajectory over a much larger number of people at the level
of countries or even the world.

Related work on \emph{metapopulation}
\cite{ColizzaV2007, ColizzaPV2007, ColizzaV2008, WangL2014}
also develops population~center-level models, 
but uses a mean~field-type approximation, which \emph{assumes}
the existence of a (sharp) threshold and finds it.
In contrast, sharp thresholds \emph{emerge} in our work.
Like \cite{GaneshMT2005, FagnaniZ2017}, we directly 
characterize the time it takes for epidemic extinction.
But unlike \cite{GaneshMT2005, FagnaniZ2017}, where there is a gap between 
the conditions for a short- and long-lasting epidemic, we prove there
is a sharp threshold for the curing rate which separates the conditions
for short- and long-lasting epidemics.%
\footnote{Note that the ``mean~field'' described in \cite{FagnaniZ2017} is 
over the network, not the infection probabilities.}
Note that our model is at the
population~center-level (compared to the person-level model in
\cite{GaneshMT2005, FagnaniZ2017}).
Besides the work on metapopulation, other
prior work which claim a sharp threshold between the two regimes 
\cite{VanMieghemOK2008, ChakrabartiWWLF2008, SahnehVMS2017, SahnehCS2012}
have assumed it and employed a mean~field-type approximation.
{Our analysis is significantly different
from the analysis of the extinction time of the mean-field
dynamics.
The advantage of this
stochastic-analysis framework is that it allows the possibility of
obtaining tail bounds for the extinction time, whereas the 
existing mean-field models, in their current form, do not offer that scope.
While we do not present tail bounds on extinction time in this work,
in Sec.~\ref{sec:numerical-computations}, we plot the confidence bounds on
the extinction times obtained from simulations of the stochastic dynamics.}

Another key aspect of our model is that the per-person  infectiousness
of the epidemic is a function of the number of active cases in the system.
{
State-dependent infectiousness influences the epidemic trajectory as
people tend to take more precautions \cite{YanMBFCO2021, FenichelKC2013, SpringbornCMF2015} 
and governments tend to impose more restrictions on travel, gatherings, etc. 
\cite{Bourassa2021, YanMBFCO2021} as the number of active cases increases.
Moreover, these changes in contact can be well-described using changes in 
the parameters of standard epidemiological compartment 
models \cite{ChangPKGRGL2020}; models that incorporate these considerations
may yield predictions that are significantly different from
models that do not \cite{FenichelCCCPHHHMPSVV2011}.
}
As explained in \cite{Sattenspiel1990}, modeling the effects 
of human behavior on epidemic spread is necessary for
realistic models.
Although time-dependent infectiousness has been studied empirically in 
\cite{ChenLCL2020}, we analytically model infectiousness as a function 
of the number
of active cases in the system, which provides a (tractable)
theoretical basis to time-varying infectiousness.
A person-level model for state-dependent infectiousness has been developed
in \cite{FagnaniZ2017}, but as explained earlier, modeling the epidemic
at the population~center-level allows us to predict the epidemic
trajectory over a much larger number of people.
We prove the population~center-level model has a sharp epidemic
threshold for the extinction times, in contrast to the gap between the
conditions for short- and long-lasting epidemics in \cite{FagnaniZ2017}.
Related to this are \cite{SahnehVMS2017} and \cite{SahnehCS2012}, which
develop person-level models where individual people get \emph{alerted} 
in the presence of infected neighbors and take more precautions or change 
their contacts.

Other related work on epidemic extinction time include \cite{Holme2013} which
estimates extinction time in SIR networks using simulations; 
\cite{Khatri2020} which calculates the extinction-time distribution in an
aggregate non-network model; \cite{HolmeT2018} which computes the mean
extinction times for all possible configurations of small networks;
\cite{HindesS2016, ChenHZL2017} which use the Wentzel-Kramers-Brillouin
approximation; and \cite{BallH2017} which characterizes the epidemic
extinction times over a ``mean'' network formed from a given degree
distribution.

To summarize, our main contribution is a sharp, analytical, and direct 
(not mean-field) characterization of
the extinction time in a population~center-level model with 
state-dependent infectiousness. 
This, to the best of our knowledge, is new.

The remainder of this paper is organized as follows.
Sec.~\ref{sec:model} describes our model.
Under this model, 
Sec.~\ref{sec:sharp-threshold} proves the existence of a sharp threshold:
if the curing rate \(\delta\) is greater than this threshold, the mean time 
for epidemic die-out starting from a state with a cumulative of \(n\) infections is of order 
\(\ln n\), and if the curing rate is below this threshold, the mean die-out
time is infinite.
Sec.~\ref{sec:extension} generalizes the results to settings with
asymmetric and weighted graphs.
Then Sec.~\ref{sec:vanishing-infectiousness} proves that the
asymptotic mean extinction time is (exactly) equal to \(\frac{\ln n}{\delta}\)
independent of graph structure if the per-person infectiousness functions 
go to zero asymptotically.
This would happen if the level of precautions people take to
combat the epidemic keep getting more stringent with increasing numbers
of active cases.
Sec.~\ref{sec:numerical-computations} provides simulation and computation
results, and Sec.~\ref{sec:conclusion} concludes.

\section{Model}
\label{sec:model}

Let there be a set of \emph{localities}%
\footnote{Note that ``localities'' can refer to
population centers at various levels of demographic aggregation.
They could represent countries, states, cities, or even
neighborhoods within a city. 
Indeed, there can be marked differences in how
people react to a contagion even within a single large urban area
\cite{KimCCJL2017}.
}
\(\mathcal{L}\), and at each locality
\(u \in \mathcal{L}\), the number of infected people at time
\(t\) is given by \(X_u(t)\).
We assume each locality has a large enough population that for our
purposes, for all \(u\), the range of \(X_u(t)\) is the set of 
all non-negative integers.
There is a graph \(\mathcal{G}\) across the localities, and \((u, v) \in \mathcal{G}\)
when the localities \(u\) and \(v\) are \emph{connected}.
The adjacency matrix \(G\) of \(\mathcal{G}\) is the matrix
having \(G_{uv}=1\) if \((u,v)\in\mathcal{G}\) and \(G_{uv}=0\) otherwise.
For ease of presentation, we first assume that the graph is symmetric:
\((u,v)\in\mathcal{G}\) implies \((v,u)\in\mathcal{G}\).
We relax this assumption in Sec.~\ref{sec:extension}.
A connection between two localities means that infected people in one
locality can infect susceptible people in the other locality.
Further, we assume the graph \(\mathcal{G}\) is connected,
i.e., for every \(u,v\in\mathcal{L}\), there exists a path between
\(u\) and \(v\) in \(\mathcal{G}\).

\begin{figure}[h]
\begin{center}
\resizebox{0.8\columnwidth}{!}{
    \begin{tikzpicture}
        \filldraw [black] (-1,0) circle (1.5pt);
        \filldraw [black] (1,0) circle (1.5pt);
        \filldraw [black] (0.3,0.6) circle (1.5pt);
        \filldraw [black] (-0.3,-0.6) circle (1.5pt);
        \begin{scope}[>=stealth, decoration={markings,mark=at position 0.25 with {\arrow[red]{<}}, mark=at position 0.75 with {\arrow[red]{>}}}]
           \draw[postaction={decorate}] (-1,0) -- (1,0);
           \draw[postaction={decorate}] (-1,0) -- node[above, red, midway, sloped] {\tiny \(\beta(X(t))\)} (0.3,0.6);
           \draw[postaction={decorate}] (1,0) -- (0.3,0.6);
           \draw[postaction={decorate}] (1,0) -- (-0.3,-0.6);
        \end{scope}
        \begin{scope}[black!30!brown, >=stealth, decoration={markings, mark=at position 1 with {\arrow{>}}}]
            \draw[postaction={decorate}] (-1,0.1) -- (-1,0.3);
            \draw[postaction={decorate}] (-0.3,-0.5) -- (-0.3,-0.3);
            \draw[postaction={decorate}] (0.3,0.7) -- (0.3,0.9);
            \draw[postaction={decorate}] (1,0.1) -- (1,0.3);
        \end{scope}
            \node at (0.2, 0.9) [anchor=west, black!30!brown, yshift=0.07cm] {\tiny \(\beta^\textsc{int}(X(t))\)};
            \draw[->, thick, black!30!green, >=stealth] (1.5, 0.5) --  node[anchor=west] {\scriptsize \(\delta\)} (1.5, -0.5);
    \end{tikzpicture}
}
\end{center}
\caption{The epidemic model, where
nodes represent population centers, edges represent 
the connections between the centers,
\(\beta(\cdot)\) and \(\beta^\textsc{int}(\cdot)\) are the
between-locality and intra-locality infectiousness functions, and \(\delta\)
is the curing rate.
}
\label{fig:model}
\end{figure}
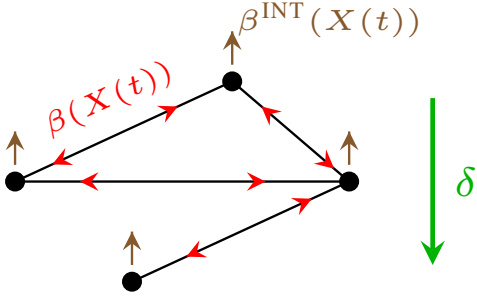

Let the total number of people infected at time \(t\) be \(X(t)\),
i.e.,  \(\sum_{u\in\mathcal{L}}X_u(t) = X(t)\).
The rate of growth of the infection at locality \(u\) at time \(t\)
consists of two components: 
\begin{enumerate}
\item the intra-locality growth rate
due to interactions within the 
locality given by \(\beta^\textsc{int}(X(t))X_u(t)\), and
\item the between-locality growth rate, where the rate of growth due to \(v\)
for each \((u,v)\in\mathcal{G}\) is given by \(\beta(X(t))X_v(t)\).
\end{enumerate}
Here, \(\beta(\cdot)\) and \(\beta^\textsc{int}(\cdot)\) are positive
real-valued functions of the total number of infections in the system,
which give the rate of
growth of the infection per infecting agent.
We assume that these per-person infectiousness functions are bounded.
Let their suprema be given by \(\sup_{n\in\N}\beta(n)=\beta_{\max}\) and
\(\sup_{n\in\N}\beta^\textsc{int}(n)=\beta^\textsc{int}_{\max}\).
We also assume that the asymptotic limits for these functions exist as
the total number of infections grows without bound: 
\(\lim_{n \to \infty}\beta(n)=\beta_\infty\) and
\(\lim_{n \to \infty}\beta^\textsc{int}(n)=\beta^\textsc{int}_\infty\).
Let the curing rate for every infected agent be \(\delta\).
This is independent of the graph \(\mathcal{G}\), the 
level of precautions taken (\(\beta\) and 
\(\beta^\textsc{int}\)), or the number of  
infections at any node \(\{X_u(t)\}\), and just depends on
the nature of the infection.
We show the model pictorially in Fig.~\ref{fig:model}.

For each \(u\in\mathcal{L}\), the above discussion implies the following rates for the infection:
\begin{align}
X_u(t) &\rightarrow X_u(t)+1\ \nonumber\\
\qquad\qquad&\text{at rate}\ \sum_{v:(u,v)\in\mathcal{G}}\!\!\!\beta(X(t))X_v(t) + \beta^{\textsc{int}}(X(t))X_u(t), \nonumber\\
X_u(t) &\rightarrow X_u(t)-1\ \text{at rate}\ \delta X_u(t).
\label{eq:node-rates}
\end{align}
Let us use the vector \(\mathbf{X}(t)\) to denote the state of the system at
time \(t\).
The \(u\)th element of \(\mathbf{X}(t)\) is \(X_u(t)\),
the number of infections at node \(u\) at time \(t\).
Let \(T_\mathbf{X}\) denote the time it takes to go from a 
state \(\mathbf{X}\) to the all-zero state \(\mathbf{0}\).
Once the epidemic reaches the all-zero state, it is \emph{extinct,} since
one can only contract the infection from someone else,%
\footnote{Note that this is true for many viral infections since the only
host for these viruses are humans. However, this may not be the case
for other infections.}
and if there are
no infected individuals, the epidemic can never rebound later.
The mean extinction time (also called the
mean \emph{hitting time}) starting from the state \(\mathbf{X}\) is given
by \(\EX\left[T_\mathbf{X}\right]\).

\section{Sharp Threshold}
\label{sec:sharp-threshold}

In this section, we state our main result as Theorem~\ref{thm:spectral-bound}.

\begin{theorem}
Let \(\lim_{n\to\infty}\beta(n)=\beta_\infty\) and 
\(\lim_{n\to\infty}\beta^\textsc{int}(n)=\beta^\textsc{int}_\infty\).
Let \(\lambda_r\) denote the spectral radius of the 
adjacency matrix of the (symmetric) undirected graph \(\mathcal{G}\).
Let the system start in some state \(\mathbf{X}\) that has \(n\)
infections cumulatively, i.e., \(\mathbf{1}^\top\mathbf{X}=n\).
If the graph \(\mathcal{G}\) is connected, then the following hold.
\begin{itemize}
\item[(i)] If \(\beta_\infty\lambda_r+\beta^\textsc{int}_\infty<\delta\),
then \(\EX\left[T_\mathbf{X}\right] \le C \ln n\)
for some constant \(C>0\).
\item[(ii)] If \(\beta_\infty\lambda_r+\beta^\textsc{int}_\infty>\delta\),
then \(\EX\left[T_\mathbf{X}\right]=\infty\).
\end{itemize}
\label{thm:spectral-bound}
\end{theorem}

Before we prove Theorem~\ref{thm:spectral-bound}, let us first observe
a property of the spectral radius, \(\lambda_r\), of \(G\).
Since we have assumed that \(\mathcal{G}\) is connected, 
the Perron-Frobenius theorem (see \cite{GraphSpectraBook})
implies that every element of the eigenvector 
\(\mathbf{q}\) corresponding to
\(\lambda_{r}\) is strictly positive, i.e., \(\mathbf{q}\succ0\).

We prove Theorem~\ref{thm:spectral-bound} in two parts:
(i) \(\beta_\infty\lambda_r+\beta^\textsc{int}_\infty<\delta\), and
(ii) \(\beta_\infty\lambda_r+\beta^\textsc{int}_\infty>\delta\).

\subsection{Curing rate above the threshold}
For proving part (i) of Theorem~\ref{thm:spectral-bound}, 
the following claim, which follows from analyzing the time evolution of
\(\mathbf{X}\) along the eigen-direction of \(G\), is useful.
\begin{claim}
When \(\beta(n)\) and \(\beta^\textsc{int}(n)\) are constant, i.e.,
\(\beta(n)=\beta\) and \(\beta^\textsc{int}(n)=\beta^\textsc{int}\) for all
\(n\), and \(\beta\lambda_r+\beta^\textsc{int}<\delta\), then
\(\EX\left[T_\mathbf{X}\right] \le C \ln n\) for some \(C>0\)
where \(\mathbf{1}^\top\mathbf{X}=n\).
\label{claim:constant-infection-spectral-lower}
\end{claim}
\begin{IEEEproof}
Please see Appendix~\ref{appendix:constant-infection-spectral-lower}.
\end{IEEEproof}

We are now ready to prove part (i) of Theorem~\ref{thm:spectral-bound}.
\begin{IEEEproof}[Proof of part {\normalfont (i)} of Theorem~\ref{thm:spectral-bound}]
Since we have \(\lim_{n\to\infty}\beta(n)=\beta_\infty\),
\(\lim_{n\to\infty}\beta^\textsc{int}(n)=\beta^\textsc{int}_\infty\),
and \(\beta_\infty\lambda_r+\beta^\textsc{int}_\infty<\delta\),
it follows from the definition of limit \cite{ThomasF1996} that
there is an \(m\) such that
\begin{align}
\beta(n)\lambda_r + \beta^\textsc{int}(n) < \delta,\ 
\text{for all}\ n \ge m.
\label{eq:ngem-lower}
\end{align}
When the system starts in any state with a cumulative number of 
infections \(n\), which is greater than \(m\), it must go through a state
where the cumulative number of infections is \(m\) to reach the all-zero
state.
However, if the system starts in a state that has less than \(m\) infections
in total, then it may or may not reach a state with \(m\) infections.
This gives us 
\begin{align*}
\EX\left[T_\mathbf{X}\right] \le \EX\left[T_{\mathbf{X},m}\right]
+ \max_{\mathbf{Y}:\mathbf{1}^{\!\!\top}\!\mathbf{Y}=m}\EX\left[T_\mathbf{Y}\right],
\end{align*}
where \(T_{\mathbf{X},m}\) is the amount of time it takes to reach a state
with a total of \(m\) infections starting from state \(\mathbf{X}\).

Using \eqref{eq:ngem-lower}, we make the following observations.
\begin{enumerate}
\item[(a)] Between \(\mathbf{X}\) and any state with a total of \(m\)
infections, \emph{every} state satisfies
\(\beta(\cdot)\lambda_r+\beta^\textsc{int}(\cdot)<\delta\), and thus a 
system with a constant infectiousness equal to 
\(\max_{n\ge m}\left(\beta(n)\lambda_r+\beta^\textsc{int}(n)\right)\)
satisfies Claim~\ref{claim:constant-infection-spectral-lower}.
\item[(b)] Since our system has an infectiousness less than the system with
constant infectiousness in point~(a) for \emph{every} state with \(n\ge m\),
using a stochastic-dominance argument,
the time it takes for epidemic extinction in the constant-infectiousness
system should be greater (in expectation) than the time our system takes to
go from \(\mathbf{X}\) to a state with \(m\) infections.
\end{enumerate}
Using these observations and Claim~\ref{claim:constant-infection-spectral-lower},
it follows that
\(\EX\left[T_{\mathbf{X},m}\right] \le C \ln n\) whenever \(n \ge m\).
Since \(\max_{\mathbf{Y}:\mathbf{1}^{\!\!\top}\!\mathbf{Y}=m}\EX\left[T_\mathbf{Y}\right]\) is a constant independent of \(n\),
we have
\begin{align*}
\EX\left[T_\mathbf{X}\right] \le C' \ln n,
\end{align*}
when \(\mathbf{1}^\top\mathbf{X}=n\) and \(n \ge m\).
For \(n<m\), \(\EX\left[T_\mathbf{X}\right]\) is less than the constant
\(\max_{\mathbf{Y}:\mathbf{1}^{\!\!\top}\!\mathbf{Y}=m}\EX\left[T_\mathbf{Y}\right]\),
and hence \(\EX\left[T_\mathbf{X}\right] \le C' \ln n\) follows directly.
\end{IEEEproof}

\subsection{Curing rate below the threshold}

We now move to the case where \(\beta_\infty\lambda_r+\beta^\textsc{int}_\infty>\delta\).
For this case, we use the discrete-time Markov chain (DTMC)
embedded in 
the continuous-time Markov chain (CTMC) \(\mathbf{X}(t)\).
Let \(\mathbf{X}_0=\mathbf{X}(0)\), and let \(\mathbf{X}_k\) be the state
of our system after \(k\) transitions.
Then \(\mathbf{X}_0, \mathbf{X}_1, \ldots\) form a DTMC.
However, the number of transitions in the CTMC 
must be countable for every sample path of the CTMC if the
embedded DTMC is to include every transition in the CTMC.
If the transitions in the CTMC are otherwise uncountably infinite,
we cannot map all the transitions in the CTMC to transitions in the DTMC.

If the CTMC's transitions are countable, and if the embedded DTMC is transient, there is a nonzero
probability that the sequence 
\(\mathbf{X}_0, \mathbf{X}_1, \mathbf{X}_2, \ldots\) 
does not contain the all-zero state \(\mathbf{0}\), with
zero infections at all nodes.
This in turn implies there is a nonzero probability that our system
does not reach the zero state starting from \(n\) infections initially
(because the transitions are countable).
This gives us an infinite mean hitting time \(\EX[T_{\mathbf{X}_0}]\).

We first state as Claim~\ref{claim:ctmc-countable} 
that our system has a countable number of transitions.
We use this together with a theorem from \cite{MalyshevM1979} (stated as
Theorem~\ref{thm:potential} here)
to prove the transience of our system
when \(\beta_\infty\lambda_r+\beta^\textsc{int}_\infty>\delta\).
\begin{claim}
Let \(\mathbb{T}\) be the set of all transition times for the continuous-time Markov chain
given by \eqref{eq:node-rates}.
There exists an injection from \(\mathbb{T}\) to \N\ with probability
\(1\), i.e., the set \(\mathbb{T}\) is countable.
\label{claim:ctmc-countable}
\end{claim}
\begin{IEEEproof}
Please see Appendix~\ref{appendix:ctmc-countable} for the proof.
Claim~\ref{claim:ctmc-countable} is similar to the results in
\cite[Section~5.1]{ResnickAdventures}.
\end{IEEEproof}

To show that the embedded discrete-time Markov chain is transient,
the following theorem from \cite{MalyshevM1979}
(paraphrased in our notation) is useful.
\begin{theorem}[from \cite{MalyshevM1979}]
Let the state space of the Markov chain 
\(\mathbf{X}_0, \mathbf{X}_1, \ldots\) be given by \(\mathcal{S}\).
If there exists a function \(V:\mathcal{S}\mapsto\R_+\cup\{0\}\) 
that satisfies
the following properties:
\begin{enumerate}
\item[(a)] for some \(d>0\), 
\(\PR\Big(\lvert V(\mathbf{X}_{k+1})-V(\mathbf{X}_k)\rvert>d\Big) = 0\ \ \text{for all}\ \ \mathbf{X}_{k}\ \text{and}\ \mathbf{X}_{k+1}\),
\item[(b)] for some \(\epsilon>0\), and \(c>0\),
\(\EX\Big[V(\mathbf{X}_{k+1})-V(\mathbf{X}_k)\mid\mathbf{X}_k=\mathbf{X}\Big]>\epsilon\ \text{for all}\ \mathbf{X}\in\{\mathbf{Y} \mid V(\mathbf{Y})\ge c\}\),
\end{enumerate}
then the Markov chain \(\mathbf{X}_0,\mathbf{X}_1,\ldots\) is transient.
\label{thm:potential}
\end{theorem}
Note that the conditions for transience in Theorem~\ref{thm:potential} are
similar to Foster's well-known work~\cite{Foster1953}.
While the conditions for positive recurrence from \cite{Foster1953} 
are still widely used, the conditions for transience require
the potential function \(V\) to be bounded.
The conditions for transience given in Theorem~\ref{thm:potential} 
from~\cite{MalyshevM1979} are easier to use.
See \cite{SrikantY2013} for other variants.

\begin{IEEEproof}[Proof of part {\normalfont (ii)} of Theorem~\ref{thm:spectral-bound}]
We first prove that the DTMC embedded in our CTMC satisfies the
conditions of Theorem~\ref{thm:potential}, which implies that the
embedded DTMC is transient.
Claim~\ref{claim:ctmc-countable} then ensures that the transience
of the embedded DTMC implies transience of the CTMC.

For the embedded DTMC, let us define the potential function 
\begin{align*}
V(\mathbf{X})=\mathbf{q}^\top\mathbf{X}.
\end{align*}
Recall that \(\mathbf{q}\) is the Perron-Frobenius eigenvector of
\(G\),
which ensures that \(\mathbf{q}\succ0\) and so \(V(\mathbf{X})\) is a valid
potential function.
This gives us \(V(\mathbf{X}_{k+1})-V(\mathbf{X}_k)=\mathbf{q}^\top\left(\mathbf{X}_{k+1}-\mathbf{X}_k\right)\).
Condition~(a) of Theorem~\ref{thm:potential} is straightforward to verify since
\(\mathbf{X}_{k+1}-\mathbf{X}_k=\pm \mathbf{e}_i\) for some \(i\),
where \(\mathbf{e}_i\) is the vector whose \(i\)th element is \(1\)
and the rest are \(0\).
So \(\PR\big(\lvert V(\mathbf{X}_{k+1})-V(\mathbf{X}_k)\rvert>d) = 0\)
for all \(d>q_{\max}\), where \(q_{\max}\) is the maximum element
of \(\mathbf{q}\).

We now define \(c\), and thus the set \(\{\mathbf{Y}\mid V(\mathbf{Y})\ge c\}\) in condition~(b) of Theorem~\ref{thm:potential}.
We set \(c=q_{\max}m\), where \(m\) shall be determined later.
This means that a sufficient condition for the transience of the
embedded DTMC is that condition~(b) of Theorem~\ref{thm:potential} should
hold in the set \(\mathcal{U}=\{\mathbf{Y}\mid V(\mathbf{Y})\ge q_{\max}m\}\).
Note that since we have defined \(V(\mathbf{Y})=\mathbf{q}^\top\mathbf{Y}\),
\(V(\mathbf{Y})\ge q_{\max}m\) implies \(\mathbf{1}^\top\mathbf{Y} \ge m\).

Let \(\mathbf{X}_k=\mathbf{X}\in\mathcal{U}\), and let
the sum of all rates in \eqref{eq:node-rates} when the system is
in this state be \(R\).
Let \(\mathbf{1}^\top\mathbf{X}=n\).
Using \eqref{eq:node-rates}, we get
\begin{align*}
\EX[V(\mathbf{X}_{k+1})-&V(\mathbf{X}_k)\mid\mathbf{X}_k=\mathbf{X}] \\
&=
\mathbf{q}^\top\EX\left[\mathbf{X}_{k+1}-\mathbf{X}_k\mid\mathbf{X}_k=\mathbf{X}\right] \\
&=
\mathbf{q}^\top\times\frac{1}{R}\Big(\beta(n) G \mathbf{X} + \beta^\textsc{int}(n)\mathbf{X} -
\delta \mathbf{X}\Big) \\
&= \frac{1}{R}(\beta(n)\lambda_{r}+\beta^\textsc{int}(n)-\delta)\mathbf{q}^\top\mathbf{X}.
\end{align*}

Observe that
\begin{align}
R &=
\mathbf{1}^\top\left(\beta(n) G \mathbf{X}+\beta^\textsc{int}(n)\mathbf{X}
+\delta\mathbf{X}\right) \nonumber \\
&\le (\beta(n) d_{\max}+\beta^\textsc{int}(n)+\delta)\mathbf{1}^\top\mathbf{X},
\label{eq:total-transition-rate}
\end{align}
where \(d_{\max}\) is the maximum degree in the graph.

Since \(\mathbf{q}^\top\mathbf{X} \ge q_{\min}\mathbf{1}^\top\mathbf{X}\),
where \(q_{\min}\) is the minimum element of \(\mathbf{q}\), we get
\begin{align*}
\EX[V(\mathbf{X}_{k+1})-V(\mathbf{X}_k)&\mid\mathbf{X}_k=\mathbf{X}] \\
&\ge \frac{(\beta(n)\lambda_{r}+\beta^\textsc{int}(n)-\delta)q_{\min}}{
\beta(n) d_{\max}+\beta^\textsc{int}(n)+\delta}
\end{align*}
for all \(\mathbf{X}\in\mathcal{U}\).
Since \(\beta(n)\to\beta_\infty\) and 
\(\beta^\textsc{int}(n)\to\beta^\textsc{int}_\infty\), 
the definition of limit ensures that for a sufficiently large 
\(m\), \(\frac{(\beta(n)\lambda_{r}+\beta^\textsc{int}(n)-\delta)q_{\min}}{\beta(n) d_{\max}+\beta^\textsc{int}(n)+\delta}\)
is arbitrarily close to 
\(\frac{(\beta_\infty\lambda_{r}+\beta^\textsc{int}_\infty-\delta)q_{\min}}{
\beta_\infty d_{\max}+\beta^\textsc{int}_\infty+\delta}\)
for all \(n \ge m\).
Since \(\frac{(\beta_\infty\lambda_{r}+\beta^\textsc{int}_\infty-\delta)q_{\min}}{
\beta_\infty d_{\max}+\beta^\textsc{int}_\infty+\delta}>0\), we have
\begin{align*}
\EX[V(\mathbf{X}_{k+1})-V(\mathbf{X}_k)&\mid\mathbf{X}_k=\mathbf{X}]
\ge \epsilon > 0
\end{align*}
for all \(\mathbf{X}\in\mathcal{U}\) for a sufficiently large \(m\).
(Recall that \(\mathcal{U}=\{\mathbf{Y}\mid V(\mathbf{Y})\ge q_{\max}m\}\)
which implies \(\mathbf{1}^\top\mathbf{X}\ge m\) for all \(\mathbf{X}\in\mathcal{U}\).)
This proves the transience of the embedded DTMC.

Transience of the embedded DTMC means that starting
in state \(\mathbf{X}_0=\mathbf{X}\,(\neq\mathbf{0})\), there is a nonzero
probability that the sequence of states 
\(\mathbf{X}_1,\mathbf{X}_2, \mathbf{X}_3, \ldots\) does not contain the
all-zero state \(\mathbf{0}\) with nonzero probability
(directly from the definition of transience used in 
\cite{MalyshevM1979} in their proof of Theorem~\ref{thm:potential}).
Using Claim~\ref{claim:ctmc-countable}, this means that the CTMC defined
in \eqref{eq:node-rates} has a nonzero probability of never reaching
the all-zero state.
Hence the average hitting time is infinite.
\end{IEEEproof}

\section{Extension to General Networks}
\label{sec:extension}

So far, we have assumed that the connection graph among the population
centers
is symmetric (\(G_{uv}=G_{vu}\)) and unweighted (\(G_{uv}\in\{0,1\}\)).
However, this is not true for many real-world networks:
the rate of infection spread between any two connected centers need not be
identical, and the rate of infection spread from \(u\) to \(v\) need not be
equal to the rate of infection spread from \(v\) to \(u\) for a connected
pair \((u,v)\).
Thus, it is important to study the behavior of the epidemic under a general
connection network given by a general asymmetric, (nonnegative) real-valued
adjacency matrix \(G\).%
\footnote{
Rather than defining the graph \(\mathcal{G}\) as a set
\(\{(u,v)\}\) of node pairs,
we now define it as a set of triples \(\{(u,v,e_{uv})\}\), where \(e_{uv}\)
is the weight of the edge from \(u\) to \(v\).
The adjacency matrix \(G\) concisely captures all this information.
}
However, it is still reasonable to assume that the graph 
is strongly connected, i.e., there exists a path with nonzero edges from any
center \(u\) to any other center \(v\).
This is because
it is rarely the case that there exist no paths from one population
center to another.

Further, the intra-locality growth rate
of infections need not be identical for all the population centers, as
this rate typically depends on local factors like 
population density \cite{WongL2020} and social capital \cite{VarshneyS2020}.
Let us use the parameter \(D_u>0\) to modulate the growth rate of the infection at location \(u\).
Let \(D\) be a diagonal matrix with \(D_u\) as the \(u\)th element of its diagonal.

These considerations give us the following expressions 
for the rates of epidemic spread.
\begin{align}
\mathbf{X}(t) &\to \mathbf{X}(t) + \mathbf{e}_u \nonumber \\
&\quad\quad\quad \text{at rate}\ 
\left[\Big(\beta(X(t))G+\beta^\textsc{int}(X(t))D\Big)\mathbf{X}(t)\right]_u,\nonumber \\
\mathbf{X}(t) &\to \mathbf{X}(t) - \mathbf{e}_u
\ \text{at rate}\
\delta\left[\mathbf{X}(t)\right]_u,
\label{eq:general-rates}
\end{align}
for all \(u\in\mathcal{L}\), where \(\left[\cdot\right]_u\) indicates the
\(u\)th element of a vector.

Let \(\rho(\cdot)\) denote the spectral radius of a matrix.
We generalize Theorem~\ref{thm:spectral-bound}
as Theorem~\ref{thm:general-bound}.
\begin{theorem}
Let the system start in some state \(\mathbf{X}\) that has \(n\)
infections cumulatively, i.e., \(\mathbf{1}^\top\mathbf{X}=n\).
For the epidemic described by \eqref{eq:general-rates}, the following hold.
\begin{itemize}
\item[(i)] If \(\rho\left(\beta_\infty G+\beta^\textsc{int}_\infty D\right)<\delta\),
then \(\EX\left[T_\mathbf{X}\right] \le C \ln n\)
for some constant \(C>0\).
\item[(ii)] If \(\rho\left(\beta_\infty G+\beta^\textsc{int}_\infty D\right)>\delta\),
then \(\EX\left[T_\mathbf{X}\right]=\infty\).
\end{itemize}
\label{thm:general-bound}
\end{theorem}
\begin{IEEEproof}
Note that the Perron-Frobenius theorem holds for the matrix 
\(\beta_\infty G + \beta^\textsc{int}_\infty D\), and we can find
a strictly positive eigenvector \(\mathbf{q}'\succ0\) of 
\(\beta_\infty G + \beta^\textsc{int}_\infty D\), which has the 
(positive, real) eigenvalue \(\rho\left(\beta_\infty G + \beta^\textsc{int}_\infty D\right)\) (see \cite{GraphSpectraBook}).
The proof follows directly by replacing the \(\mathbf{q}\) used in 
Claim~\ref{claim:constant-infection-spectral-lower} and 
Theorem~\ref{thm:spectral-bound} with the Perron-Frobenius eigenvector
of \(\beta_\infty G + \beta^\textsc{int}_\infty D\).
\end{IEEEproof}

While Theorem~\ref{thm:general-bound} provides a sharp threshold in terms
of \(\rho\left(\beta_\infty G + \beta^\textsc{int}_\infty D\right)\),
it is difficult to separate the contributions of the between-locality
spreading term \(\beta_\infty G\) and the intra-locality spreading
term \(\beta^\textsc{int}_\infty D\).
It would be nice to have sufficient conditions for fast die-out and
long-lasting epidemic in terms of expressions where these two contributions
are decoupled.
Towards this end, we provide two corollaries.

\begin{corollary}
Let the system start in some state \(\mathbf{X}\) that has \(n\)
infections cumulatively, i.e., \(\mathbf{1}^\top\mathbf{X}=n\).
If \(D\) is a scalar matrix \(\eta I\), i.e., if the intra-locality 
rate-modulating factor \(D_u=\eta\) for every locality \(u\),
then the following hold.
\begin{itemize}
\item[(i)] If \(\beta_\infty\rho(G)+\beta^\textsc{int}_\infty\eta<\delta\),
then \(\EX\left[T_\mathbf{X}\right] \le C \ln n\)
for some constant \(C>0\).
\item[(ii)] If \(\beta_\infty\rho(G)+\beta^\textsc{int}_\infty\eta>\delta\),
then \(\EX\left[T_\mathbf{X}\right]=\infty\).
\end{itemize}
\label{cor:intra-constant}
\end{corollary}
\begin{IEEEproof}
Please see Appendix~\ref{appendix:intra-constant}.
\end{IEEEproof}

For the next corollary, we need a theorem from \cite{Schwenk1986} which
relates the spectral radius of nonnegative
asymmetric matrices to the spectral radius of certain symmetric matrices.
We state this as Claim~\ref{claim:schwenk} (in a form useful for us).

\begin{claim}[from \cite{Schwenk1986}]
For any nonnegative (square) matrix \(A\),
\begin{align*}
\rho\left(\sqrt{A \odot A^\top}\right) \le \rho(A) \le \rho\left(\frac{A+A^\top}{2}\right),
\end{align*}
where \(\odot\) is the element-wise product of matrices and \(\sqrt{\cdot}\)
is the element-wise square root.
\label{claim:schwenk}
\end{claim}
Note that the \(ij\)th element of \(\sqrt{A \odot A^\top}\) is 
\(\sqrt{A_{ij}A_{ji}}\) and the \(ij\)th element of 
\(\frac{A+A^\top}{2}\) is \(\frac{A_{ij}+A_{ji}}{2}\).
Both \(\sqrt{A \odot A^\top}\) and \(\frac{A+A^\top}{2}\) are symmetric
matrices.
This reduction to symmetric matrices allows us to apply Weyl's inequalities on the conditions in
Theorem~\ref{thm:general-bound}.
We state this formally as Corollary~\ref{cor:weyl-derived}.
See the textbook \cite{Bhatia1997} for the details regarding Weyl's inequalities. 
We also provide short proofs of the inequalities used here in Appendix~\ref{appendix:weyl-proof}.

\begin{corollary}
Let the system start in some state \(\mathbf{X}\) that has \(n\)
infections cumulatively, i.e., \(\mathbf{1}^\top\mathbf{X}=n\).
Then the following hold.
\begin{itemize}
\item[(i)] If \(\beta_\infty\rho\left(\frac{G+G^\top}{2}\right)+\beta^\textsc{int}_\infty\max_uD_u<\delta\),
then \(\EX\left[T_\mathbf{X}\right] \le C \ln n\)
for some constant \(C>0\).
\item[(ii)] If \(\beta_\infty\rho\left(\sqrt{G \odot G^\top}\right)+\beta^\textsc{int}_\infty\min_uD_u>\delta\),
then \(\EX\left[T_\mathbf{X}\right]=\infty\).
\end{itemize}
\label{cor:weyl-derived}
\end{corollary}
\begin{IEEEproof}
Applying the upper bound in Claim~\ref{claim:schwenk} to the spectral-radius expression in
part (i) of Theorem~\ref{thm:general-bound}, we get
\begin{align*}
\rho\left(\beta_\infty G + \beta^\textsc{int}_\infty D\right)
\le
\rho\left(\beta_\infty\tfrac{G+G^\top}{2}+\beta^\textsc{int}_\infty D\right).
\end{align*}
Since \(\beta_\infty\frac{G+G^\top}{2}\) and \(\beta^\textsc{int}_\infty D\)
are both symmetric matrices, we can apply one of Weyl's inequalities  
(see \cite{Bhatia1997} or Appendix~\ref{appendix:weyl-proof}) to get
\begin{align}
\rho\left(\beta_\infty G + \beta^\textsc{int}_\infty D\right)
\le
\beta_\infty \rho\left(\tfrac{G+G^\top}{2}\right) + \beta^\textsc{int}_\infty \max_u D_u.
\label{eq:weyl-upper-in-proof}
\end{align}
Equation~\eqref{eq:weyl-upper-in-proof} ensures that whenever the condition in part (i) of Corollary~\ref{cor:weyl-derived} is satisfied, the condition in part (i) of Theorem~\ref{thm:general-bound} is satisfied as well.
This proves part (i) of Corollary~\ref{cor:weyl-derived}.

For part (ii) of Corollary~\ref{cor:weyl-derived}, observe that
\begin{align*}
(\beta_\infty G + \beta^\textsc{int}_\infty D)&\odot
(\beta_\infty G + \beta^\textsc{int}_\infty D)^\top \\
&= \beta_\infty^2 G \odot G^\top + (\beta^\textsc{int}_\infty)^2 D \odot D.
\end{align*}
This is because there is no position \(ij\) that has a nonzero element in
both the matrices \(G\) and \(D\).
Further, the matrix \(D\) is diagonal (and hence symmetric), and so we have
\begin{align*}
&\sqrt{
(\beta_\infty G + \beta^\textsc{int}_\infty D)\odot
(\beta_\infty G + \beta^\textsc{int}_\infty D)^\top
} \\
&\qquad\qquad\qquad\qquad
= \beta_\infty\sqrt{G \odot G^\top} + \beta^\textsc{int}_\infty D.
\end{align*}
Using the lower bound in Claim~\ref{claim:schwenk}, we get
\begin{align*}
\rho\left(\beta_\infty G + \beta^\textsc{int}_\infty D\right)
\ge
\rho\left(\beta_\infty\sqrt{G \odot G^\top}+\beta^\textsc{int}_\infty D\right).
\end{align*}
Since \(\beta_\infty\sqrt{G \odot G^\top}\) and \(\beta^\textsc{int}_\infty D\) are both symmetric matrices, we can apply another one of Weyl's
inequalities (see \cite{Bhatia1997} or Appendix~\ref{appendix:weyl-proof})
to get 
\begin{align*}
\rho\left(\beta_\infty G + \beta^\textsc{int}_\infty D\right)
\ge \beta_\infty\rho\left(\sqrt{G \odot G^\top}\right) + \beta^\textsc{int}_\infty \min_u D_u.
\end{align*}
Thus, whenever the condition in part (ii) of Corollary~\ref{cor:weyl-derived} is true, the condition in part (ii) of Theorem~\ref{thm:general-bound} is true as well.
This concludes the proof of part (ii) of Corollary~\ref{cor:weyl-derived}.
\end{IEEEproof}

Unlike Theorem~\ref{thm:spectral-bound}, Theorem~\ref{thm:general-bound},
and Corollary~\ref{cor:intra-constant} where the thresholds are sharp,
there is a gap between the thresholds for a quick die-out
and long-lasting epidemic in Corollary~\ref{cor:weyl-derived}.
However, Corollary~\ref{cor:weyl-derived} decouples the contributions
of the graph structure \(G\) and the variation in intra-locality spreading 
\(D\) in the thresholds.

\section{Vanishing Infectiousness}
\label{sec:vanishing-infectiousness}

In this section, we consider the special case where the per-person
infectiousness functions decrease to zero as the number of active cases
in the system increases:
\(\beta_\infty=\beta^\textsc{int}_\infty=0\).
For this, we define upper-bound and lower-bound Markov chains
using the maximum and minimum node degrees.
We then show that both these Markov chains have the same asymptotic
mean hitting times if the per-person infectiousness functions go to
zero asymptotically.

Let the maximum node in-degree
in \(\mathcal{G}\) be \(d_{\max}\) and the 
minimum node in-degree be \(d_{\min}\).%
\footnote{
For weighted graphs, use the definitions
\(d_{\max}=\max_u\sum_vG_{uv}\) and \(d_{\min}=\min_u\sum_vG_{uv}\).
}
Adding up \eqref{eq:node-rates} over all the localities \(u\in\mathcal{L}\) gives us the following 
\emph{upper-} and \emph{lower-bound} Markov chains for the system-wide epidemic.
\begin{align}
&\textit{Upper-bound Markov chain:} \nonumber\\
&X(t) \rightarrow X(t)+1\ \text{at rate}\ \Big(d_{\max}\beta\big(X(t)\big)\!\!+\!\! \beta^\textsc{int}\big(X(t)\big)\Big)X(t), \nonumber \\
&X(t) \rightarrow X(t)-1\ \text{at rate}\ \delta X(t),
\label{eq:ub-rates}
\end{align}
and
\begin{align}
&\textit{lower-bound Markov chain:} \nonumber\\
&X(t) \rightarrow X(t)+1\ \text{at rate}\ \Big(d_{\min}\beta\big(X(t)\big)\!\!+\!\! \beta^\textsc{int}\big(X(t)\big)\Big)X(t), \nonumber \\ 
&X(t) \rightarrow X(t)-1\ \text{at rate}\ \delta X(t).
\label{eq:lb-rates}
\end{align}
The mean hitting times of these upper- and lower-bound Markov chains are, 
respectively, 
higher and lower than the mean hitting times of the original epidemic.
Proofs that they are in fact bounds are straightforward.

We can see that the form of both \eqref{eq:ub-rates} for the
upper-bound Markov chain and \eqref{eq:lb-rates} for the
lower-bound Markov chain can be captured using a rate coefficient
\(\gamma(\cdot)\) as follows.
\begin{align}
&X(t) \to X(t)+1\ \text{at rate}\ \gamma(X(t))X(t), \nonumber\\
&X(t) \to X(t)-1\ \text{at rate}\ \delta X(t).
\label{eq:rates}
\end{align}
Any results we derive for a general \(\gamma(\cdot)\) apply for both the
upper-bound and lower-bound Markov chains.
So we now derive bounds for the hitting times of a general Markov chain
satisfying \eqref{eq:rates}.

Let \(T_n\) be  the time it takes for the infection to go to \(0\) infections starting
from \(n\) infections.
Starting from \(n\) infections, 
the probability that the system given by \eqref{eq:rates} goes to \(n+1\) infections next
(instead of \(n-1\) infections) is given by 
\(\frac{\gamma(n)}{\gamma(n)+\delta}\).
Similarly, the probability that the system goes to \(n-1\) infections
next after \(n\) infections is given by \(\frac{\delta}{\gamma(n)+\delta}\).
This gives us
\begin{align*}
\EX[T_n] &=
\EX[T_{n+1}]\frac{\gamma(n)}{\gamma(n)+\delta} + \EX[T_{n-1}]\frac{\delta}{\gamma(n)+\delta} + \EX[\tau_n],
\end{align*}
where \(\tau_n\) is the time it takes to make the next transition from \(n\) infections.
Using \(\EX[\tau_n]=\frac{1}{n(\gamma(n)+\delta)}\), rearranging the terms,
and replacing
\(n\) with \(n-1\) throughout, we get
\begin{align*}
\EX[T_{n}] &= \EX[T_{n-1}]\tfrac{\gamma(n-1)+\delta}{\gamma(n-1)} - \EX[T_{n-2}]\tfrac{\delta}{\gamma(n-1)}
- \tfrac{1}{(n-1)\gamma(n-1)}
\end{align*}
for \(n \ge 2\).
Defining \(S_n = \EX[T_n]-\EX[T_{n-1}]\) yields
\begin{align}
S_{n+1}\gamma(n) - S_{n}\delta = -\frac{1}{n}
\label{eq:sn-recursion}
\end{align}
for \(n \ge 1\) with \(S_1=\EX[T_1]\).

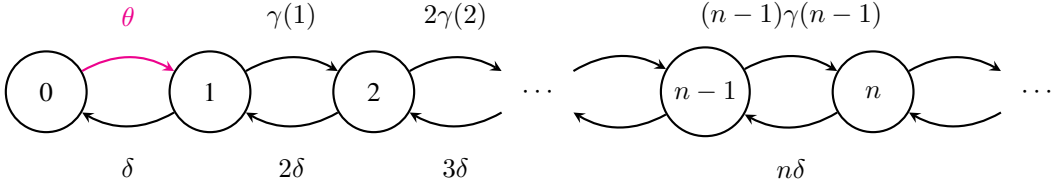
\begin{figure*}[t!]
\begin{center}
\resizebox{0.8\textwidth}{!}{
    \begin{tikzpicture}[>=stealth,thick]
        \tikzset{minimum size=3em}
        \node (0) [circle, draw] {0};
        \node (1) [right=3em of 0, circle, draw] {1};
        \node (2) [right=3em of 1, circle, draw] {2};
        \node (dot) [right=3em of 2, circle] {\(\cdots\)};
        \path (0)
            edge[bend left, ->, magenta] node [above] 
                                    {\textcolor{magenta}{\(\theta\)}} (1)(1)
            edge[bend left, ->] node [above] {\(\gamma(1)\)} (2)(2)
            edge[bend left, ->] node [above] {\(2\gamma(2)\)} (dot);
        \path (dot)
            edge[bend left, ->] node [pos=0.5, below] {\(3\delta\)} (2)(2)
            edge[bend left, ->] node [below] {\(2\delta\)} (1)(1)
            edge[bend left, ->] node [below] {\(\delta\)} (0);
        \node (n) [right=3em of dot, circle, draw] {\(n-1\)};
        \node (n1) [right=3em of n, circle, draw] {\(n\)};
        \node (lastdot) [right=3em of n1, circle] {\(\cdots\)};
        \path (dot)
            edge[bend left, ->] node [above] {} (n)(n)
            edge[bend left, ->] node [above] {\((n-1)\gamma(n-1)\)} (n1)(n1)
            edge[bend left, ->] node [above] {} (lastdot);
        \path (lastdot)
            edge[bend left, ->] node [below] {} (n1)(n1)
            edge[bend left, ->] node [below] {\(n\delta\)} (n)(n)
            edge[bend left, ->] node [below] {} (dot);
    \end{tikzpicture}
}
\caption{Modified Markov chain with same mean hitting times as the Markov
chain in \eqref{eq:rates}.
Adding \(\theta\) does not change the hitting times, but makes the chain
positive-recurrent.}
\label{fig:modified-markov}
\end{center}
\end{figure*}

So if we can find \(\EX[T_1]\), we will be able to compute all the 
mean hitting times (not necessarily in closed form).
To compute \(\EX[T_1]\), we compute the steady-state probability in 
state \(0\) of
the transformed Markov chain in Fig.~\ref{fig:modified-markov},
whose hitting times are the same as the required Markov chain in \eqref{eq:rates}.
The modification we have done to the Markov chain in \eqref{eq:rates}
is the addition of the extra transition out of the zero state with a rate
\(\theta\).
This does not change the hitting time from any nonzero state since the
time it takes to reach the zero state for the first time is
independent of the rate of transition out of the zero state.
However, the transformation gives us a positive-recurrent Markov chain, 
for which the steady-state probabilities are well-defined and non-trivial.
Further, the mean hitting times are independent of the birth rate from \(0\), \(\theta\).

Let \(\pi_n\) be the steady-state probability of finding the chain in node \(n\).
Local balance between node \(n-1\) and node \(n\) gives
\begin{align*}
\pi_{n-1}(n-1)\gamma(n-1) = \pi_nn\delta,
\end{align*}
which on expanding out yields
\begin{align*}
\pi_n &= 
\theta\pi_0\frac{\gamma(1)\gamma(2)\cdots\gamma(n-1)}{n\delta^n},
\end{align*}
for \(n \ge 1\).
Using \(\sum_{n=0}^\infty\pi_n=1\), we get
\begin{align}
\pi_0\left(
1 + \theta\left(
\frac{1}{\delta} + \frac{\gamma(1)}{2\delta^2} + \frac{\gamma(1)\gamma(2)}{3\delta^3} +
\cdots 
\right)
\right)
=1.
\label{eq:pi0}
\end{align}
From renewal theory (see \cite[Chapter~7]{Ross2019}), we have
\begin{align*}
\pi_0 = \frac{\EX[\tau_0]}{\EX[\tau_0]  + \EX[T_1]}.
\end{align*}
Since the rate of transition out of the zero state 
(in the modified Markov chain) is \(\theta\), 
\(\EX[\tau_0]=\frac{1}{\theta}\), and this  gives
\begin{align*}
\EX[T_1] = \frac{1}{\theta}\left(\frac{1}{\pi_0}-1\right).
\end{align*}
Substituting the expression for \(\pi_0\) from \eqref{eq:pi0} implies the following claim.
\begin{claim}
\label{claim:t1-gamma}
The mean hitting time from one infected agent to zero infected agents is given by
\begin{align*}
{
    \EX[T_1] = \frac{1}{\delta}\sum_{i=1}^\infty \frac{1}{i}\frac{\prod_{j=1}^{i-1}\gamma(j)}{\delta^{i-1}}
}
\end{align*}
whenever the Markov chain in Fig.~\ref{fig:modified-markov} is positive
recurrent.
\end{claim}

Our goal in this section has been to compute the 
asymptotic mean hitting times
when \(\beta_\infty\) and \(\beta^\textsc{int}_\infty\) are \(0\).
These conditions translate to \(\lim_{n\to\infty}\gamma(n)=0\) for both
the upper-bound Markov chain \eqref{eq:ub-rates} and the lower-bound
Markov chain \eqref{eq:lb-rates}.
We get there by first computing the (asymptotic) mean hitting times when
\(\gamma(n)=\alpha\), which we do in the next subsection.

\subsection{Hitting time bounds when \(\gamma(\cdot)\) is a constant}

Substituting \(\gamma(n)=\alpha\) in the expression for \(\EX[T_1]\) in
Claim~\ref{claim:t1-gamma}, we get
\begin{align}
{
\EX[T_1] = \frac{1}{\delta}\sum_{i=1}^\infty \frac{1}{i}\left(\frac{\alpha}{\delta}\right)^{i-1}
}
\label{eq:t1-alpha}
\end{align}
and expanding out \eqref{eq:sn-recursion} for \(\gamma(n)=\alpha\)
gives us
\begin{align*}
S_n = S_{n-1}\frac{\delta}{\alpha} - \frac{1}{\alpha(n-1)}\qquad\quad \\
\end{align*}
\vspace{-3em}
{
\begin{align*}
\qquad\qquad&= \frac{\delta^2}{\alpha^2}S_{n-2} - \frac{\delta}{\alpha^2(n-2)} - \frac{1}{\alpha(n-1)} \\
&\ \ \vdots \\
&= \frac{\delta^{n-1}}{\alpha^{n-1}}\left(
S_1 - \frac{1}{\delta}\sum_{i=1}^{n-1}\frac{1}{i}\left(\frac{\alpha}{\delta}\right)^{i-1}
\right)
\end{align*}
}
Since \(S_1=\EX[T_1]\) by definition, substituting the expression from
\eqref{eq:t1-alpha} gives us
{
\begin{align*}
S_n &= 
\frac{\delta^{n-1}}{\alpha^{n-1}}\cdot\frac{1}{\delta}\sum_{i=n}^\infty\frac{1}{i}\left(\frac{\alpha}{\delta}\right)^{i-1} \\
&= \frac{1}{\delta n}\sum_{i=0}^\infty\frac{n}{n+i}\left(\frac{\alpha}{\delta}\right)^{i},
\end{align*}
}
and since \(\frac{n}{n+r}<1\) for all positive integers \(r\), we get
\begin{align*}
\frac{1}{\delta n}\ \le\ &S_n\ \le\ \frac{1}{(\delta-\alpha)n},
\end{align*}
using the geometric series \(1+\frac{\alpha}{\delta}+\frac{\alpha^2}{\delta^2}+\cdots=\frac{\delta}{\delta-\alpha}\),
which implies
\begin{align*}
\frac{1}{\delta}\sum_{i=1}^n\frac{1}{i}\ \le\ &\EX[T_n]\ \le\ \frac{1}{\delta-\alpha}\sum_{i=1}^n\frac{1}{i}.
\end{align*}
This directly leads us  to the following claim.
\begin{claim}
When the per-person infectiousness is given by \(\gamma(n)=\alpha\) for 
all \(n\) for some
\(\alpha\in(0,\delta)\), the mean hitting time to go 
to zero infections starting from \(n\) infections satisfies
\begin{align*}
\frac{\ln (n+1)}{\delta} \ \ \le\ \ \EX[T_n]\ \ \le\ \ \frac{1 + \ln n}{\delta-\alpha}.
\end{align*}
\label{claim:tn-constant}
\end{claim}

\subsection{When \(\lim_{n\to\infty}\gamma(n)=0\)}

When the infectiousness functions \(\beta(\cdot)\) 
and \(\beta^\textsc{int}(\cdot)\) go to zero, i.e., 
\(\beta_\infty=0\) and \(\beta^\textsc{int}_\infty=0\),
the \(\gamma(\cdot)\) for both the upper-bound Markov chain in
\eqref{eq:ub-rates} and the 
lower-bound Markov chain in \eqref{eq:lb-rates} go to zero.
Hence, if we can derive the asymptotic mean hitting time for
\(\lim_{n\to\infty}\gamma(n)=0\), it will give us matching asymptotes
for the upper and lower bounds, 
which means we have the exact asymptote.

We will show that for any arbitrarily small \(\alpha\), we can use 
Claim~\ref{claim:tn-constant} to show that the asymptote for \(\EX[T_n]\) is
arbitrarily close to \(\frac{\ln n}{\delta}\).
We state this formally as Theorem~\ref{thm:log-hitting}.
\begin{theorem}
If \(\lim_{n\to\infty}\gamma(n)=0\), then the mean hitting times of 
the Markov chain in Fig.~\ref{fig:modified-markov} satisfy
\begin{align*}
\lim_{n\to\infty}\frac{\delta\EX[T_n]}{\ln n} = 1.
\end{align*}
\label{thm:log-hitting}
\end{theorem}

Before proving Theorem~\ref{thm:log-hitting}, let us first state a
claim which will be useful.
\begin{claim}
If \(\lim_{n\to\infty}\gamma(n)=0\), then for any \(\epsilon > 0\),
the Markov chain  in Fig.~\ref{fig:modified-markov} satisfies
\begin{align*}
\frac{\ln (n+1)}{\delta} \le \EX[T_n] \le \frac{\ln n}{\delta - \epsilon} + h(\epsilon)\quad \text{for all}\ n,
\end{align*}
for some function \(h(\epsilon)\) that is independent of \(n\).
\label{claim:epsilon-bound}
\end{claim}
\begin{IEEEproof}
Please see Appendix~\ref{appendix:epsilon-bound}.
\end{IEEEproof}

We are now ready to prove Theorem~\ref{thm:log-hitting}.
\begin{IEEEproof}[Proof of Theorem~\ref{thm:log-hitting}]
Proving \(\lim_{n\to\infty}\frac{\delta\EX[T_n]}{\ln n}=1\) is equivalent
to proving that for any \(\epsilon > 0\), we can find an \(n_\epsilon\)
such that \(\left|\frac{\delta\EX[T_n]}{\ln n}-1\right| < \epsilon\) for
all \(n > n_\epsilon\) (from the definition of limit \cite{ThomasF1996}).

For any \(\epsilon\), substitute 
\(\min\left(\frac{\epsilon\delta}{4}, \frac{\delta}{2}\right)\) for
\(\epsilon\) in Claim~\ref{claim:epsilon-bound}.
This gives us
\begin{align*}
\frac{\delta\EX[T_n]}{\ln n} - 1 \le \frac{\epsilon}{2} + 
\frac{\delta\max\left(h\left(\frac{\epsilon\delta}{4}\right),h\left(\frac{\delta}{2}\right)\right)}{\ln n}.
\end{align*}
For sufficiently large \(n\), we get
\begin{align*}
\frac{\delta\EX[T_n]}{\ln n} - 1 < \epsilon.
\end{align*}

Further, from the lower bound in Claim~\ref{claim:epsilon-bound}, we get
\begin{align*}
\frac{\delta\EX[T_n]}{\ln n}-1 \ge \frac{\ln (n+1)}{\ln n}-1.
\end{align*}
For a sufficiently large \(n\), \(\frac{\ln(n+1)}{\ln n}-1\) 
can be made arbitrarily close to \(0\).
Thus we get
\begin{align*}
\left|\frac{\delta\EX[T_n]}{\ln n}-1\right| < \epsilon
\end{align*}
for all sufficiently large \(n\), which concludes the proof.
\end{IEEEproof}

\subsection{Putting it together for the original epidemic on \(\mathcal{G}\)}

For both the upper-bound Markov chain in \eqref{eq:ub-rates} and 
the lower-bound Markov chain in \eqref{eq:lb-rates}, the
infectiousness per person goes to zero if both \(\beta(\cdot)\) and
\(\beta^\textsc{int}(\cdot)\) go to zero as \(n \to \infty\).
Since Theorem~\ref{thm:log-hitting} applies for any chain with 
\(\lim_{n\to\infty}\gamma(n)=0\), both these upper- and lower-bound
Markov chains satisfy Theorem~\ref{thm:log-hitting}.
Since both these chains have the same asymptote, by sandwiching,
even the original epidemic
on \(\mathcal{G}\) must have the same asymptote.
This gives us the following corollary.
\begin{corollary}
If \(\lim_{n\to\infty}\beta(n)=0\) and \(\lim_{n\to\infty}\beta^\textsc{int}(n)=0\), then for any locality graph \(\mathcal{G}\), we have
\begin{align*}
\lim_{n \to \infty}\frac{\delta\EX[T_n]}{\ln n} = 1,
\end{align*}
where \(T_n\) is the time taken by the epidemic to go from a cumulative 
of \(n\) infections in the system to \(0\).
\label{cor:graph-hitting}
\end{corollary}
Corollary~\ref{cor:graph-hitting} implies that if the per-person
infectiousness functions go to zero asymptotically, i.e., if the 
(non-pharmaceutical) precautions get arbitrarily more stringent as the
number of cases increases, then the mean hitting times have the asymptote
\(\frac{\ln n}{\delta}\) independent of the locality graph.

\section{Simulations \& Numerical Computations}
\label{sec:numerical-computations}

In this section, we  present some simulations and numerical
computations to demonstrate the theoretical results of the preceding
sections.

\subsection{Network-wide simulations}

For simulations, we use the network from \cite{ColizzaPV2007}
which is a graph where the nodes represent the top \(500\) US airports and the
edge weights are the number of seats scheduled on flights between the airports
in the year 2002.
We consider the top \(100\) of these \(500\) nodes and normalize the
adjacency matrix with the mean column weight
(this normalization just scales the values of \(\beta(\cdot)\)).
We simulate the model described in Sec.~\ref{sec:model} using Gillespie's
algorithm \cite{Gillespie1977}.

\begin{center}
\begin{figure}
\subfloat[Curing rate below the threshold.]{\label{fig:graph-constant:below}
\includegraphics[width=\columnwidth]{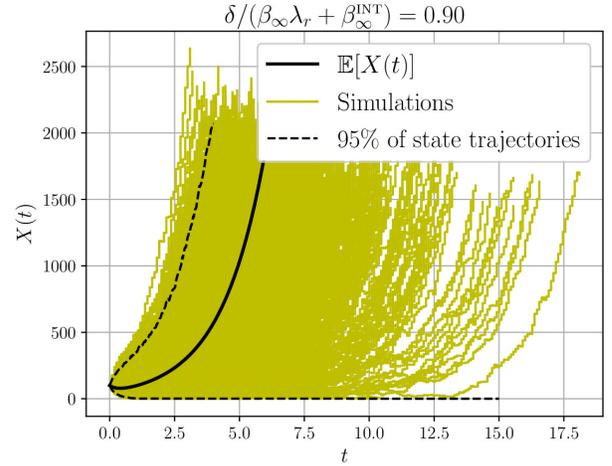}}

\subfloat[Curing rate above the threshold.]{\label{fig:graph-constant:above}
\includegraphics[width=\columnwidth]{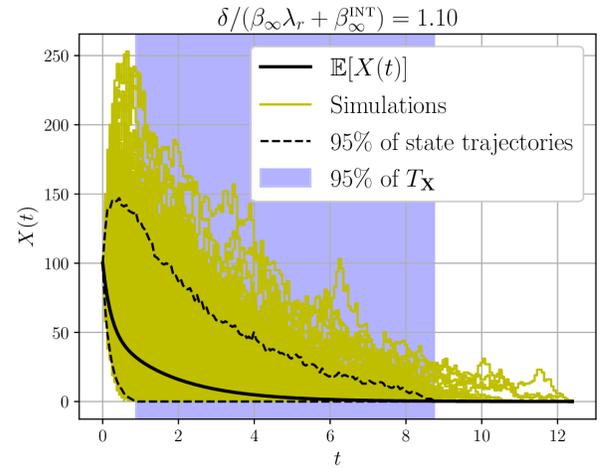}}
\caption{Epidemic trajectories using constant values for \(\beta(\cdot)\) and \(\beta^\textsc{int}(\cdot)\).}
\label{fig:graph-constant}
\end{figure}
\end{center}

First, in Fig.~\ref{fig:graph-constant}, we simulate using constant
values for \(\beta(\cdot)\) and \(\beta^\textsc{int}(\cdot)\).
Specifically, we set \(\beta(n)=\beta=2\) and
\(\beta^\textsc{int}(n)=\beta^\textsc{int}=2\) for all \(n\),
and choose \(\delta\) to get the value of 
\(\frac{\delta}{\beta\lambda_r+\beta^\textsc{int}}\) shown
on the plot.
For both the values of \(\delta\), we simulate the system {\(1000\)} times
and show the trajectories of \(X(t)\)  over time in the plot,
{and the interval that contains \(95\%\) of the simulated
states at each time instant.
We obtain this \(95\%\) interval by finding the maximum and minimum
state values after ignoring the top and
bottom \(2.5\%\) of the simulations.
}%
We also show the plot of \(\EX[X(t)]\) computed theoretically by solving
the differential equation for \(\frac{d\EX[X(t)]}{dt}\) 
(see Appendix~\ref{appendix:constant-infection-spectral-lower}).
{
As we can see in Fig.~\ref{fig:graph-constant:below}, when
\(\beta\lambda_r+\beta^\textsc{int}>\delta\), most of the simulated
trajectories of the system show an epidemic that is not dying out.
Even though more than \(2.5\%\) of the simulations die out (as the
\(95\%\) interval shows), since most of the simulations show an
epidemic that becomes increasingly larger with time, the expected
extinction time would be infinite, in line with what we have
theoretically proven in Theorem~\ref{thm:spectral-bound}.
On the other hand, in Fig.~\ref{fig:graph-constant:above},
when \(\beta\lambda_r+\beta^\textsc{int}<\delta\),
all the trajectories of the system result in the epidemic dying out
relatively quickly.
Further, in this case, the confidence bounds on the extinction time
are meaningfully defined, and
we show the \(95\%\) confidence interval of the extinction time
\(T_\mathbf{X}\) in Fig.~\ref{fig:graph-constant:above}.
This interval is calculated in the same way as the \(95\%\)
interval for the state trajectory.
}%
For all the simulations, we start with an initial epidemic size  of 
\(100\), placed uniformly at random at one of the nodes.

\begin{center}
\begin{figure}
\subfloat[Curing rate is greater than threshold when epidemic is
small.]{
\includegraphics[width=\columnwidth]{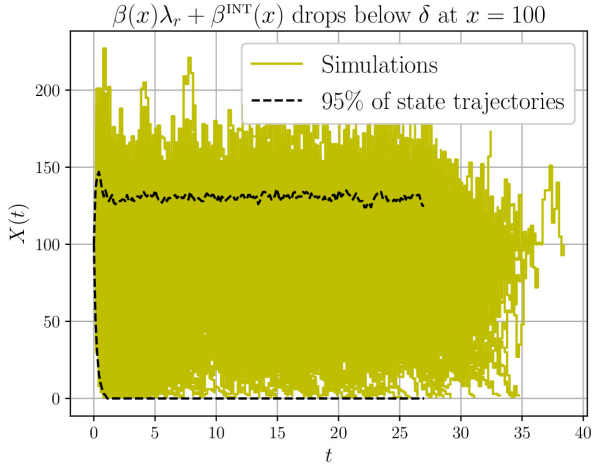}
}

\subfloat[Curing rate is greater than threshold only after epidemic
gets very large.]{
\includegraphics[width=\columnwidth]{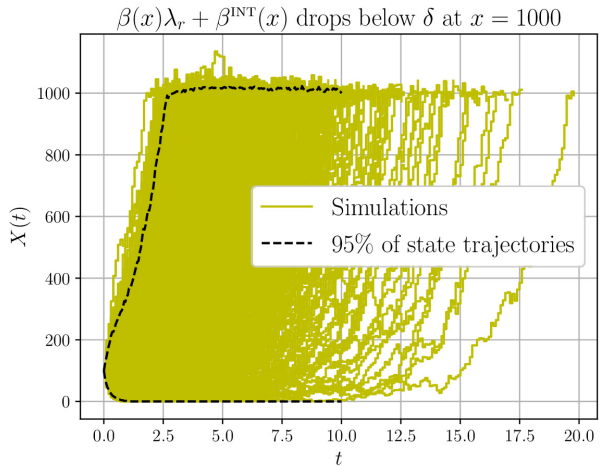}
}
\caption{Epidemic trajectories when \(\beta(n)\) and \(\beta^\textsc{int}(n)\) change with \(n\).}
\label{fig:graph-variable-drop}
\end{figure}
\end{center}

When the values of \(\beta(n)\) and \(\beta^\textsc{int}(n)\) change with
\(n\), if \(\beta(n)\lambda_r+\beta^\textsc{int}(n)<\delta\) 
or \(\beta(n)\lambda_r+\beta^\textsc{int}(n)>\delta\) for all
\(n\), then the results are very similar to the ones in Fig.~\ref{fig:graph-constant}, and hence we omit these plots.
In Fig.~\ref{fig:graph-variable-drop}, we show the results of simulations
where \(\beta(n)\lambda_r+\beta^\textsc{int}(n)\) starts from a value
greater than \(\delta\) for small \(n\), but eventually falls to a value
smaller than \(\delta\) for larger \(n\).
The value of \(n\) where this transition happens is shown on the plots
in Fig.~\ref{fig:graph-variable-drop}.
We can see in Fig.~\ref{fig:graph-variable-drop} that there seems to be
a ``metastable'' state at the point 
where the infectiousness is equal to the curing rate.
Note that since the value of \(\beta(n)\lambda_r+\beta^\textsc{int}(n)\)
eventually falls below \(\delta\) for large enough \(n\), the
condition in part (i) of Theorem~\ref{thm:spectral-bound} is true,
and so the mean hitting time should be logarithmic in the initial
infection size.
However, these simulations suggest that the epidemic takes a very  long time
to die out in this case.
It seems that the die-out  times are in fact exponential
in the infection size where the infectiousness and curing rate are equal.
Please see Appendix~\ref{appendix:exponential-equal} for some insight
into this behavior.
This means that even though Theorem~\ref{thm:spectral-bound} guarantees
that the mean die-out time would be logarithmic in the initial infection size
if the asymptotic rate of infectiousness is less than the curing rate,
it is still very important that measures such as lockdowns and other
non-pharmaceutical precautions are implemented in the early stages
of an epidemic.

\subsection{Numerical computations for  vanishing \(\gamma(\cdot)\)}

\begin{figure}
\begin{center}
\includegraphics[width=0.9\columnwidth]{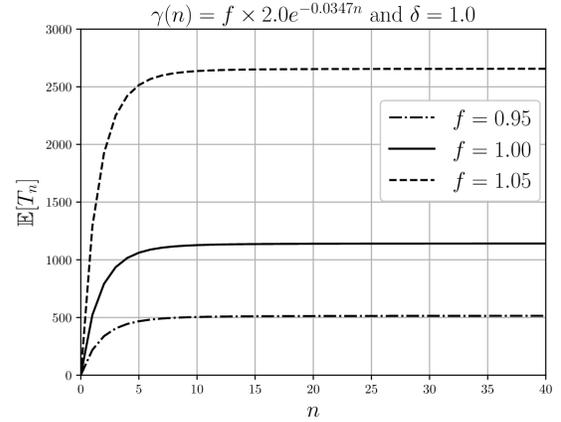}
\caption{\(\EX[T_n]\) for slightly different \(\gamma(\cdot)\) functions.}
\label{fig:multiplicative-factor}
\end{center}
\end{figure}

Here,  we provide some numerical computations to support
Theorem~\ref{thm:log-hitting}.
Note that in contrast to the network-wide simulations in
Fig.~\ref{fig:graph-constant} and \ref{fig:graph-variable-drop} 
where we have used
the infectiousness functions \(\beta(\cdot)\) and \(\beta^\textsc{int}(\cdot)\), 
we use \(\gamma(\cdot)\) here which captures the infectiousness for
both the upper- and lower-bound Markov chains together in a single 
expression using \eqref{eq:rates}.
We consider three different \(\gamma(\cdot)\)
functions and plot the values of \(\EX[T_n]\) computed using the
recursion from \eqref{eq:sn-recursion} 
(with the base case from Claim~\ref{claim:t1-gamma}).
We plot this in Fig.~\ref{fig:multiplicative-factor}.

Fig.~\ref{fig:multiplicative-factor} shows that even small changes
in \(\gamma(\cdot)\) can cause large changes in the values of \(\EX[T_n]\).
Further, Fig.~\ref{fig:multiplicative-factor} may seem to indicate that even
these small changes cause the mean hitting times to not converge to the same
asymptote.
This would be contrary to what we expect  
from Theorem~\ref{thm:log-hitting}.
However, the reason we do not see all the three curves in 
Fig.~\ref{fig:multiplicative-factor} converge to the same asymptote is that
the convergence happens extremely slowly.
This is not very surprising, given that the asymptote is the 
function \(\frac{\ln n}{\delta}\).
Since the logarithmic function increases very slowly, differences between
\(\EX[T_n]\) for different \(\gamma(\cdot)\) functions at small
values of \(n\) take a very long time to become insignificant, and the
\(\EX[T_n]\) values become close to each other only at very large values
of \(n\).

To demonstrate this, consider \(\gamma(n)=\frac{k}{n}\).
We choose this function because it leads to easier analysis.
Similar arguments hold for any other function as well.
Substituting this into Claim~\ref{claim:t1-gamma} gives us
\begin{align}
\EX[T_1] &= \frac{1}{\delta} + \frac{k}{1\cdot2\cdot\delta^2} +
\frac{k^2}{1\cdot2\cdot3\cdot\delta^3} + \cdots \nonumber \\
&= \frac{1}{k}\left(\frac{k}{\delta}+\frac{k^2}{2!\delta^2}+\frac{k^3}{3!\delta^3} + \cdots\right) \nonumber \\
&= \frac{e^{k/\delta}-1}{k}.
\label{eq:t1-kbyn}
\end{align}
Equation~\eqref{eq:t1-kbyn} is quite sensitive to the value of \(k\).
For example, with \(\delta=1\), we get a derivative of \(\frac{4e^5+1}{25}\approx23.79\) at \(k=5\).
Small changes in the value of \(k\) can significantly change the value 
of \(\EX[T_1]\).
We can use the recursion from \eqref{eq:sn-recursion} to analytically
find the value of \(\EX[T_2]\) to find that \(\EX[T_2]\) is even more
sensitive to the value of \(k\).
Since \(\EX[T_n]\) is of the form \(\EX[T_2]+\sum_{i=3}^nS_i\), and
\(S_n\) asymptotically reaches \(\frac{1}{n}\), these differences in 
\(\EX[T_2]\) become negligible only for a very large value of \(n\).

\begin{figure}
\begin{center}
\includegraphics[width=0.9\columnwidth]{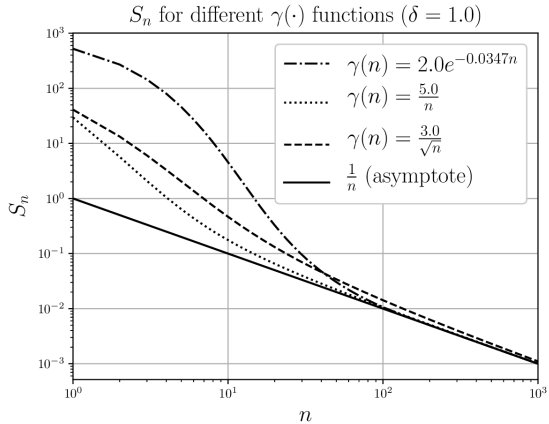}
\caption{\(S_n\) for large \(n\).
Note that for these large values of \(n\), \(\gamma(n)\) is
too small for accurately computing the recursion in 
\eqref{eq:sn-recursion} using
even 128-bit floating point arithmetic. 
To compute \(S_n\), we need to divide a very small value, \(S_{n-1}\delta-\frac{1}{n-1}\), by another very small value, \(\gamma(n-1)\).
Insufficient numerical precision can lead to garbage values for \(S_n\).
Arbitrary-precision arithmetic (such as the one provided by
\texttt{mpmath} \cite{mpmath}) is needed.}
\label{fig:difference}
\end{center}
\end{figure}

We can verify this using Fig.~\ref{fig:difference} where we plot the values
of \(S_n\) for different \(\gamma(\cdot)\) functions.
We see that all of them eventually reach the asymptote \(\frac{1}{n}\).
This means that for large enough \(n\), the mean hitting times will all be indistinguishable
from \(\ln n\).
However, we need an extremely large value of \(n\) for the differences to
become negligible.

\section{Conclusion}
\label{sec:conclusion}

We have developed a model for epidemic spread within and across
population centers with state-dependent infectiousness.
In this model,  we directly prove (without mean-field assumptions) that
there exists a sharp threshold for the curing rate \(\delta\)
such that when \(\delta\) is more than a threshold, the epidemic
dies out quickly (the mean lifetime is of logarithmic order in the initial
infection size), and when \(\delta\) is less than the threshold, the
mean lifetime of the epidemic is infinite.
Although \(\delta\) is not typically something we can control, especially
in the initial stages of a pandemic without vaccines or other medication,
it is possible to lower the threshold by following more stringent 
precautions.
While we do not provide prescriptive solutions for managing pandemics,
we hope that this work would offer useful insights to policymakers.

{

While our model makes no mean-field assumptions to
characterize the extinction time,
we provide theoretical results only on its expected value.
It is of interest to establish high-probability bounds on
extinction time
and
characterize how strongly extinction time concentrates.
Combining techniques in Claim~\ref{claim:ctmc-countable} with
literature on (discrete-time) Markov concentration 
\cite{Kotzing2016, Paulin2015} might be pursued.

There is also scope for developing broader and
more realistic models of 
state-dependent infectiousness.
Empirical work suggests
that people take precautions against contagions not only in response to
the actual number of infections, but also to other
factors like the media attention on infection prevalence \cite{FenichelKC2013, SpringbornCMF2015}.
These models should capture infectiousness as a function of both
the actual infection prevalence 
and the spread of awareness through (social) media.

Finally, it is important to accurately infer parameters of our model
using historical and current epidemiological data so as to inform
practical applications.
}

\bibliographystyle{IEEEtran}
\bibliography{venues-abrv, log-hitting}

\begin{thebibliography}{10}
\providecommand{\url}[1]{#1}
\csname url@samestyle\endcsname
\providecommand{\newblock}{\relax}
\providecommand{\bibinfo}[2]{#2}
\providecommand{\BIBentrySTDinterwordspacing}{\spaceskip=0pt\relax}
\providecommand{\BIBentryALTinterwordstretchfactor}{4}
\providecommand{\BIBentryALTinterwordspacing}{\spaceskip=\fontdimen2\font plus
\BIBentryALTinterwordstretchfactor\fontdimen3\font minus
  \fontdimen4\font\relax}
\providecommand{\BIBforeignlanguage}[2]{{%
\expandafter\ifx\csname l@#1\endcsname\relax
\typeout{** WARNING: IEEEtran.bst: No hyphenation pattern has been}%
\typeout{** loaded for the language `#1'. Using the pattern for}%
\typeout{** the default language instead.}%
\else
\language=\csname l@#1\endcsname
\fi
#2}}
\providecommand{\BIBdecl}{\relax}
\BIBdecl

\bibitem{MorensF2013}
D.~M. Morens and A.~S. Fauci, ``Emerging infectious diseases: Threats to human
  health and global stability,'' \emph{PLoS Pathog.}, vol.~9, no.~7, p.
  e1003467, Jul. 2013.

\bibitem{ShirleyR2005}
M.~D. Shirley and S.~P. Rushton, ``The impacts of network topology on disease
  spread,'' \emph{Ecol. Complex.}, vol.~2, no.~3, pp. 287--299, Sep. 2005.

\bibitem{Newman2002}
M.~E. Newman, ``Spread of epidemic disease on networks,'' \emph{Phys. Rev. E},
  vol.~66, no.~1, p. 016128, Jul. 2002.

\bibitem{GiordanoBBCDDC2020}
G.~Giordano, F.~Blanchini, R.~Bruno, P.~Colaneri, A.~Di~Filippo, A.~Di~Matteo,
  and M.~Colaneri, ``Modelling the {COVID-19} epidemic and implementation of
  population-wide interventions in {I}taly,'' \emph{Nature Med.}, vol.~26,
  no.~6, pp. 855--860, Jun. 2020.

\bibitem{ColizzaBBV2006}
V.~Colizza, A.~Barrat, M.~Barth{\'e}lemy, and A.~Vespignani, ``The modeling of
  global epidemics: Stochastic dynamics and predictability,'' \emph{Bull. Math.
  Biol.}, vol.~68, no.~8, pp. 1893--1921, Nov. 2006.

\bibitem{KuchlerRS2020}
T.~Kuchler, D.~Russel, and J.~Stroebel, ``{JUE} {I}nsight: The geographic
  spread of {COVID}-19 correlates with the structure of social networks as
  measured by {F}acebook,'' \emph{J. Urban Econ.}, p. 103314, Jan. 2021.

\bibitem{ZhouWZSYHZOPS2020}
Y.~Zhou, L.~Wang, L.~Zhang, L.~Shi, K.~Yang, J.~He, B.~Zhao, W.~Overton,
  S.~Purkayastha, and P.~Song, ``A spatiotemporal epidemiological prediction
  model to inform county-level {COVID}-19 risk in the {U}nited {S}tates,''
  \emph{Harvard Data Sci. Rev.}, Aug. 2020.

\bibitem{GomezABMM2010}
S.~G{\'o}mez, A.~Arenas, J.~Borge-Holthoefer, S.~Meloni, and Y.~Moreno,
  ``Discrete-time {M}arkov chain approach to contact-based disease spreading in
  complex networks,'' \emph{Europhys. Lett.}, vol.~89, no.~3, p. 38009, Feb.
  2010.

\bibitem{GaneshMT2005}
A.~Ganesh, L.~Massouli{\'e}, and D.~Towsley, ``The effect of network topology
  on the spread of epidemics,'' in \emph{Proc. 24th Annu. Joint Conf. IEEE
  Comput. Commun. Soc. (INFOCOM 2005)}, vol.~2, Mar. 2005, pp. 1455--1466.

\bibitem{FagnaniZ2017}
F.~Fagnani and L.~Zino, ``Diffusion of innovation in large scale graphs,''
  \emph{IEEE Trans. Netw. Sci. Eng.}, vol.~4, no.~2, pp. 100--111, 2017.

\bibitem{VanMieghemOK2008}
P.~Van~Mieghem, J.~Omic, and R.~Kooij, ``Virus spread in networks,''
  \emph{IEEE/ACM Trans. Netw.}, vol.~17, no.~1, pp. 1--14, Jun. 2008.

\bibitem{SahnehVMS2017}
F.~D. Sahneh, A.~Vajdi, J.~Melander, and C.~M. Scoglio, ``Contact adaption
  during epidemics: A multilayer network formulation approach,'' \emph{IEEE
  Trans. Netw. Sci. Eng.}, vol.~6, no.~1, pp. 16--30, Nov. 2017.

\bibitem{SahnehCS2012}
F.~D. Sahneh, F.~N. Chowdhury, and C.~M. Scoglio, ``On the existence of a
  threshold for preventive behavioral responses to suppress epidemic
  spreading,'' \emph{Sci. Rep.}, vol.~2, no. 632, Sep. 2012.

\bibitem{ColizzaV2007}
V.~Colizza and A.~Vespignani, ``Invasion threshold in heterogeneous
  metapopulation networks,'' \emph{Phys. Rev. Lett.}, vol.~99, no.~14, p.
  148701, 2007.

\bibitem{ColizzaPV2007}
V.~Colizza, R.~Pastor-Satorras, and A.~Vespignani, ``Reaction--diffusion
  processes and metapopulation models in heterogeneous networks,'' \emph{Nature
  Phys.}, vol.~3, no.~4, pp. 276--282, Apr. 2007.

\bibitem{ColizzaV2008}
V.~Colizza and A.~Vespignani, ``Epidemic modeling in metapopulation systems
  with heterogeneous coupling pattern: Theory and simulations,'' \emph{J.
  Theor. Biol.}, vol. 251, no.~3, pp. 450--467, 2008.

\bibitem{WangL2014}
L.~Wang and X.~Li, ``Spatial epidemiology of networked metapopulation: An
  overview,'' \emph{Chin. Sci. Bull.}, vol.~59, no.~28, pp. 3511--3522, 2014.

\bibitem{ChakrabartiWWLF2008}
D.~Chakrabarti, Y.~Wang, C.~Wang, J.~Leskovec, and C.~Faloutsos, ``Epidemic
  thresholds in real networks,'' \emph{ACM Trans. Inf. Syst. Secur.}, vol.~10,
  no.~4, pp. 1--26, Jan. 2008.

\bibitem{YanMBFCO2021}
Y.~Yan, A.~A. Malik, J.~Bayham, E.~P. Fenichel, C.~Couzens, and S.~B. Omer,
  ``Measuring voluntary and policy-induced social distancing behavior during
  the {COVID}-19 pandemic,'' \emph{Proc. Natl. Acad. Sci.}, vol. 118, no.~16,
  Apr. 2021.

\bibitem{FenichelKC2013}
E.~P. Fenichel, N.~V. Kuminoff, and G.~Chowell, ``Skip the trip: Air travelers'
  behavioral responses to pandemic influenza,'' \emph{PloS ONE}, vol.~8, no.~3,
  p. e58249, Mar. 2013.

\bibitem{SpringbornCMF2015}
M.~Springborn, G.~Chowell, M.~MacLachlan, and E.~P. Fenichel, ``Accounting for
  behavioral responses during a flu epidemic using home television viewing,''
  \emph{BMC Infectious Diseases}, vol.~15, no.~1, pp. 1--14, Dec. 2015.

\bibitem{Bourassa2021}
K.~J. Bourassa, ``State-level stay-at-home orders and objectively measured
  movement in the {U}nited {S}tates during the {COVID}-19 pandemic,''
  \emph{Psychosomatic Med.}, Dec. 2020.

\bibitem{ChangPKGRGL2020}
S.~Chang, E.~Pierson, P.~W. Koh, J.~Gerardin, B.~Redbird, D.~Grusky, and
  J.~Leskovec, ``Mobility network models of {COVID-19} explain inequities and
  inform reopening,'' \emph{Nature}, vol. 589, pp. 82--87, Jan. 2021.

\bibitem{FenichelCCCPHHHMPSVV2011}
E.~P. Fenichel, C.~Castillo-Chavez, M.~G. Ceddia, G.~Chowell, P.~A.~G. Parra,
  G.~J. Hickling, G.~Holloway, R.~Horan, B.~Morin, C.~Perrings, M.~Springborn,
  L.~Velazquez, and C.~Villalobos, ``Adaptive human behavior in epidemiological
  models,'' \emph{Proc. Natl. Acad. Sci.}, vol. 108, no.~15, pp. 6306--6311,
  Apr. 2011.

\bibitem{Sattenspiel1990}
L.~Sattenspiel, ``Modeling the spread of infectious disease in human
  populations,'' \emph{Am. J. Phys. Anthropol.}, vol.~33, no. S11, pp.
  245--276, 1990.

\bibitem{ChenLCL2020}
Y.-C. Chen, P.-E. Lu, C.-S. Chang, and T.-H. Liu, ``A time-dependent {SIR}
  model for {COVID}-19 with undetectable infected persons,'' \emph{IEEE Trans.
  Netw. Sci. Eng.}, vol.~7, no.~4, pp. 3279--3294, Sep. 2020.

\bibitem{Holme2013}
P.~Holme, ``Extinction times of epidemic outbreaks in networks,'' \emph{PLoS
  ONE}, vol.~8, no.~12, p. e84429, Dec. 2013.

\bibitem{Khatri2020}
B.~S. Khatri, ``Stochastic extinction of epidemics: how long would it take for
  {SARS-CoV}-2 to die out without herd immunity?'' \emph{medRxiv}, Aug. 2020.

\bibitem{HolmeT2018}
P.~Holme and L.~Tupikina, ``Epidemic extinction in networks: insights from the
  {\(12110\)} smallest graphs,'' \emph{New J. Phys.}, vol.~20, no.~11, p.
  113042, Nov. 2018.

\bibitem{HindesS2016}
J.~Hindes and I.~B. Schwartz, ``Epidemic extinction and control in
  heterogeneous networks,'' \emph{Phys. Rev. Lett.}, vol. 117, no.~2, p.
  028302, Jul. 2016.

\bibitem{ChenHZL2017}
H.~Chen, F.~Huang, H.~Zhang, and G.~Li, ``Epidemic extinction in a generalized
  susceptible-infected-susceptible model,'' \emph{J. Stat. Mech.}, vol. 2017,
  no.~1, p. 013204, Jan. 2017.

\bibitem{BallH2017}
F.~Ball and T.~House, ``Heterogeneous network epidemics: real-time growth,
  variance and extinction of infection,'' \emph{J. Math. Biol.}, vol.~75,
  no.~3, pp. 577--619, Sep. 2017.

\bibitem{KimCCJL2017}
C.~Kim, S.~H. Cheon, K.~Choi, C.-H. Joh, and H.-J. Lee, ``Exposure to fear:
  Changes in travel behavior during {MERS} outbreak in {S}eoul,'' \emph{KSCE J.
  Civ. Eng.}, vol.~21, no.~7, pp. 2888--2895, Nov. 2017.

\bibitem{GraphSpectraBook}
A.~E. Brouwer and W.~H. Haemers, \emph{Spectra of Graphs}.\hskip 1em plus 0.5em
  minus 0.4em\relax Springer Science \& Business Media, 2011.

\bibitem{ThomasF1996}
G.~B. {Thomas, Jr.} and R.~L. Finney, \emph{{C}alculus and {A}nalytic
  {G}eometry}, 9th~ed.\hskip 1em plus 0.5em minus 0.4em\relax Addison-Wesley
  Publishing Company, 1996.

\bibitem{MalyshevM1979}
V.~A. Maly\v{s}ev and M.~V. Men'\v{s}ikov, ``Ergodicity, continuity and
  analyticity of countable markov chains,'' \emph{Trudy Moskovskogo
  Matematicheskogo Obshchestva}, vol.~39, pp. 3--48, 1979, {E}nglish Transl. in
  \textit{Trans. Moscow Math. Soc.,} 1981.

\bibitem{ResnickAdventures}
S.~I. Resnick, \emph{Adventures in Stochastic Processes}.\hskip 1em plus 0.5em
  minus 0.4em\relax Springer Science \& Business Media, 1992.

\bibitem{Foster1953}
F.~G. Foster, ``On the stochastic matrices associated with certain queuing
  processes,'' \emph{Ann. Math. Stat.}, vol.~24, no.~3, pp. 355--360, Sep.
  1953.

\bibitem{SrikantY2013}
R.~Srikant and L.~Ying, \emph{Communication Networks: an Optimization, Control,
  and Stochastic Networks Perspective}.\hskip 1em plus 0.5em minus 0.4em\relax
  Cambridge, United Kingdom: Cambridge University Press, 2013.

\bibitem{WongL2020}
D.~W.~S. Wong and Y.~Li, ``Spreading of {COVID}-19: Density matters,''
  \emph{PLoS ONE}, vol.~15, no.~12, p. e0242398, Dec. 2020.

\bibitem{VarshneyS2020}
L.~R. Varshney and R.~Socher, ``{COVID}-19 growth rate decreases with social
  capital,'' \emph{medRxiv}, Apr. 2020.

\bibitem{Schwenk1986}
A.~J. Schwenk, ``Tight bounds on the spectral radius of asymmetric nonnegative
  matrices,'' \emph{Linear Algebra Appl.}, vol.~75, pp. 257--265, Mar. 1986.

\bibitem{Bhatia1997}
R.~Bhatia, \emph{Matrix {A}nalysis}.\hskip 1em plus 0.5em minus 0.4em\relax New
  York: Springer, 1997.

\bibitem{Ross2019}
S.~M. Ross, \emph{Introduction to Probability Models}.\hskip 1em plus 0.5em
  minus 0.4em\relax London, United Kingdom: Academic Press, 2019.

\bibitem{Gillespie1977}
D.~T. Gillespie, ``Exact stochastic simulation of coupled chemical reactions,''
  \emph{J. Phys. Chem.}, vol.~81, no.~25, pp. 2340--2361, Dec. 1977.

\bibitem{mpmath}
\BIBentryALTinterwordspacing
F.~Johansson \emph{et~al.}, \emph{mpmath: a {P}ython library for
  arbitrary-precision floating-point arithmetic (version 1.0.0)}, Sep. 2017.
  [Online]. Available: \url{http://mpmath.org/}
\BIBentrySTDinterwordspacing

\bibitem{Kotzing2016}
T.~K{\"o}tzing, ``Concentration of first hitting times under additive drift,''
  \emph{Algorithmica}, vol.~75, no.~3, pp. 490--506, Jul. 2016.

\bibitem{Paulin2015}
D.~Paulin, ``Concentration inequalities for {M}arkov chains by {M}arton
  couplings and spectral methods,'' \emph{Elec. J. Prob.}, vol.~20, pp. 1--32,
  Jan. 2015.

\end{thebibliography}

\begin{IEEEbiographynophoto}{Akhil~Bhimaraju}
received the B.Tech. and M.Tech. degrees in electrical
engineering from the Indian Institute of Technology Madras
in 2020.
He is currently a Ph.D. student with the department of
electrical and computer engineering and the
Coordinated Science Laboratory at the
University of Illinois Urbana--Champaign.
His research interests include mathematical
modeling of engineering problems.
\end{IEEEbiographynophoto}
\begin{IEEEbiographynophoto}{Avhishek~Chatterjee}
received the Ph.D. degree in electrical and computer
engineering from The University of Texas at Austin in 2015. 
From 2015 to 2017, he was a Post--Doctoral Research Associate with the 
Coordinated Science Laboratory, 
University of Illinois Urbana--Champaign. 
He is currently
an Assistant Professor with the Department of Electrical Engineering, 
IIT Madras. 
His research interests lie in theoretical studies of dynamics, optimal
designs, and operations of stochastic networks.
\end{IEEEbiographynophoto}
\begin{IEEEbiographynophoto}{Lav~R.~Varshney}
(S'00--M'10--SM'15) received the B.S. degree (\emph{magna cum laude})
with honors in electrical and computer engineering from Cornell
University, Ithaca, NY, USA, in 2004, and the S.M., E.E., and Ph.D. degrees in
electrical engineering and computer science from the Massachusetts Institute
of Technology, Cambridge, MA, USA, in 2006, 2008, and 2010, respectively.
He is currently an associate professor of electrical and computer engineering, 
with further affiliations in computer science, industrial engineering, 
neuroscience, digital agriculture, and personalized nutrition with the 
University of Illinois Urbana--Champaign.
He is also a computational scientist with Brookhaven National Laboratory, Upton, NY, USA.
\end{IEEEbiographynophoto}

\appendices

\section{Proof of Claim~\ref{claim:constant-infection-spectral-lower}}
\label{appendix:constant-infection-spectral-lower}

The following proof that we provide here closely resembles 
the proof of \cite[Theorem 3.1]{GaneshMT2005}.
However, since we are  not interested in the exact constant \(C\) like
\cite{GaneshMT2005}, we avoid the use of matrix exponentials
seen there.
The rates of \eqref{eq:node-rates} (with constant \(\beta(\cdot)\) and
\(\beta^\textsc{int}(\cdot)\)) give us the following differential
equation:
\begin{align*}
\frac{d\EX\left[\mathbf{X}(t)\right]}{dt}=
\Big(
\beta G + \beta^\textsc{int}I - \delta I
\Big)\EX\left[\mathbf{X}(t)\right],
\end{align*}
where \(I\) is the identity matrix (of correct size).
Multiply each side of the equation with \(\mathbf{q}^\top\) 
(\(\mathbf{q}\) is the eigenvector of \(G\) corresponding to
\(\lambda_r\)).
This gives us
\begin{align}
\frac{d\EX\left[\mathbf{q}^\top\mathbf{X}(t)\right]}{dt}=
\mathbf{q}^\top\Big(\beta G + \beta^\textsc{int}I - \delta I\Big)\EX[\mathbf{X}(t)].
\label{eq:preceding-equation}
\end{align}
Since \(\mathbf{q}\) is an eigenvector of \(G\) with 
eigenvalue \(\lambda_{r}\), and an eigenvector 
of \(I\) with eigenvalue \(1\) (every vector is an eigenvector of \(I\)
with eigenvalue \(1\)),
\eqref{eq:preceding-equation} gives us
\begin{align*}
\frac{d\EX\left[\mathbf{q}^\top\mathbf{X}(t)\right]}{dt}=
\Big(\beta \lambda_{r} + \beta^\textsc{int} - \delta \Big)
\EX\left[\mathbf{q}^\top\mathbf{X}(t)\right].
\end{align*}
This is a differential equation in 
terms of \(\EX\left[\mathbf{q}^\top\mathbf{X}(t)\right]\),
and solving it gives us
\begin{align*}
\EX\left[\mathbf{q}^\top\mathbf{X}(t)\right] = e^{t\left(\beta\lambda_{r}+\beta^\textsc{int}-\delta\right)}\mathbf{q}^\top\mathbf{X}(0).
\end{align*}
Let \(q_{\max}\) and \(q_{\min}\) denote the maximum and minimum elements
of \(\mathbf{q}\), i.e., \(q_{\max}=\max_{i}{q}_i\) and 
\(q_{\min}=\min_{i}{q}_i\).
Since \(\mathbf{q}\succ0\), \(q_{\min}\) is strictly positive.
This gives us
\begin{align}
\EX[X(t)] = \EX\left[\mathbf{1}^\top\mathbf{X}(t)\right] \le 
e^{t\left(\beta\lambda_{r}+\beta^\textsc{int}-\delta\right)}\frac{q_{\max}n}{q_{\min}}.
\label{eq:xt-exponential-bound}
\end{align}

The mean hitting time can be written as 
\begin{align*}
\EX\left[T_{\mathbf{X}(0)}\right] &= 
\int_{0}^\infty \PR(X(t) > 1) dt \\
&= \int_0^\tau \PR(X(t)>1) dt + \int_\tau^\infty \PR(X(t)>1) dt \\
&\le \tau + \int_\tau^\infty \EX[X(t)] dt
\end{align*}
for any \(\tau>0\).
The last inequality follows from the fact that \(\PR(X(t)>1)\le1\) since it
is a probability (which gives the first term), and  the Markov inequality
which gives us \(\PR(X(t)>1)\le\EX[X(t)]\) (for the second term).

Using \eqref{eq:xt-exponential-bound}, we get 
\begin{align*}
\EX[T_{\mathbf{X}(0)}] \le \tau+kne^{-\tau\Delta}
\quad \text{for all}\quad \tau > 0,
\end{align*}
where \(k=\frac{q_{\max}}{q_{\min}(\delta-\beta\lambda_{r}-\beta^\textsc{int})}>0\) and \(\Delta=\delta-\beta\lambda_{r}-\beta^\textsc{int}>0\).
Setting \(\tau=\frac{\ln n}{\Delta}\) gives us \(\EX[T_{\mathbf{X}(0)}] \le C \ln n\).

\section{Proof of Claim~\ref{claim:ctmc-countable}}
\label{appendix:ctmc-countable}

Divide the time axis into intervals of unit length.
Given any (finite) \(t\in\mathbb{T}\), if the number of transitions in
all intervals preceding and including \(t\) is finite, then 
the cardinality of the set \(\{s \mid s\in\mathbb{T}\ \text{and}\ s<t\}\) 
is finite.
Further, this cardinality is unique for each \(t\),
allowing us to map \(t\) to this unique natural number plus one.
Thus we get an injective mapping (if the number of transitions in each interval
is finite).

At the start of the interval, assume that the Markov chain starts 
in state \(\mathbf{X}\) with \(\mathbf{1}^\top\mathbf{X}=n\).
The probability that there are at least \(k\) transitions in the interval
satisfies
\begin{align}
\PR(\text{at least \(k\) transitions in interval})
\le
\PR\left(\sum_{j=0}^{k-1}X_j \le 1\right),
\label{eq:ktransitionsle1}
\end{align}
where \(\{X_j\}\) are the amounts of time it takes to transition out
of the first \(k\) states starting from \(\mathbf{X}\) at the beginning of the
interval.

Since the total rate of transition rate out of \(\mathbf{X}\) is given by
\(\mathbf{1}^\top\left(\beta(n)\mathbf{G}+\beta^\textsc{int}(n)\mathbf{I}+\delta\mathbf{I}\right)\mathbf{X}\),
the total transition rate out of any state with at most \(n\) infections
is less than or equal to \((\beta_{\max}d_{\max}+\beta^\textsc{int}_{\max}+\delta)n\).
Recall that \(\beta_{\max}=\sup_{i\in\N}\beta(i)\),
\(\beta^\textsc{int}_{\max}=\sup_{i\in\N}\beta^\textsc{int}(i)\),
and \(d_{\max}\) is the maximum degree among nodes of \(\mathcal{G}\).
Define \(\tau=\beta_{\max}d_{\max}+\beta^\textsc{int}_{\max}+\delta\).

So in the worst case, which gives the greatest probability on the right
side of \eqref{eq:ktransitionsle1}, we have \(X_j\sim\exp(\tau(n+j))\).
This gives us
\begin{align*}
\PR\left(\sum_{j=0}^{k-1}X_j \le 1\right)
&\le
\PR\left(e^{-\sum_{j=0}^{k-1}X_j}\ge e^{-1}\right) \\
&\le e\prod_{j=0}^{k-1}\EX\left[e^{-X_j}\right] \\
&= e\prod_{j=0}^{k-1}\frac{\tau n+\tau j}{1+\tau n+\tau j} \\
&=\frac{e}{\prod_{j=0}^{k-1}\left(1+\frac{1}{\tau n+\tau j}\right)}.
\end{align*}
If \(\prod_{j=0}^{k-1}\left(1+\frac{1}{\tau n+\tau j}\right)\to\infty\)
as \(k\to\infty\), then the probability that there are infinite transitions
in the interval goes to \(0\).
But this is equivalent to \(\sum_{j=0}^{k-1}\ln\left(1+\frac{1}{\tau n+\tau j}\right)\to \infty\)
as \(k\to\infty\).

This gives us
\begin{align*}
\sum_{j=0}^{k-1}\ln\left(1+\frac{1}{\tau n+\tau j}\right) 
&=
\sum_{j=0}^{k-1}\frac{\ln \left(1+\frac{1}{\tau n+\tau j}\right)}{\frac{1}{\tau n+\tau j}}\frac{1}{\tau n+\tau j}.
\end{align*}
For a large enough \(j\), we can make \(\frac{\ln\left(1+\frac{1}{\tau n+\tau j}\right)}{\frac{1}{\tau n+\tau j}}\) arbitrarily close to \(1\).
This implies
\begin{align*}
\sum_{j=0}^{k-1}\frac{\ln\left(1+\frac{1}{\tau n+\tau j}\right)}{\frac{1}{\tau n+\tau j}}\!\frac{1}{\tau n+\tau j}\!\! >\!\! \left(1-\epsilon\right)\!\!\sum_{j=l}^{k-1}\frac{1}{\tau n+\tau j}\to\infty,
\end{align*}
where \(l\) is chosen to be large enough so that \(\frac{\ln\left(1+\frac{1}{\tau n+\tau j}\right)}{\frac{1}{\tau n+\tau j}}\) is at most \(\epsilon\) away from
\(1\) for all \(j\ge l\).
The sum goes to infinity because the sum of the harmonic series 
goes to infinity.
Since this ensures that the Markov chain only has a finite number of
transitions in any interval, it concludes the proof.

\section{Proof of Corollary~\ref{cor:intra-constant}}
\label{appendix:intra-constant}

We need to show that
\begin{align*}
\rho\left(\beta_\infty G + \beta^\textsc{int}_\infty \eta I\right)
= \beta_\infty \rho(G) + \beta^\textsc{int}_\infty\eta.
\end{align*}
Recall that the spectral radius of a matrix is defined as the maximum
absolute value of the eigenvalues of the matrix.
Let \(\lambda\) be an eigenvalue of \(\beta_\infty G + \beta^\textsc{int}_\infty\eta I\).
This yields
\begin{align*}
\text{det}(\beta_\infty G + \beta^\textsc{int}_\infty \eta I - \lambda I)=0,
\end{align*}
or
\begin{align*}
\text{det}(\beta_\infty G - (\lambda-\beta^\textsc{int}_\infty\eta)I)=0.
\end{align*}
This implies \(\lambda-\beta^\textsc{int}_\infty\eta\) is an eigenvalue
of \(\beta_\infty G\) for every eigenvalue \(\lambda\) of \(\beta_\infty G + \beta^\textsc{int}_\infty\eta I\).
The Perron-Frobenius theorem (see \cite{GraphSpectraBook}) guarantees that there
exists a positive eigenvalue of \(\beta_\infty G\) which has the 
maximum absolute value.
Thus the maximum absolute value of \(\lambda\) is \(\beta_\infty\rho(G)+\beta^\textsc{int}_\infty\eta\).

\section{Spectral Radius of Sum of Symmetric and Diagonal Matrices}
\label{appendix:weyl-proof}

In this appendix, we prove a special case of Weyl's inequality which
suffices for the purposes of this paper.
We state this formally in Claim~\ref{claim:weyl-special}.
\begin{claim}
Let \(P\) be any nonnegative symmetric matrix and 
\(Q\) be any nonnegative diagonal matrix.
Then
\begin{align*}
\rho(P) + \min_i Q_{ii} \le \rho(P+Q) \le \rho(P) + \max_i Q_{ii},
\end{align*}
where \(\rho(\cdot)\) denotes the spectral radius.
\label{claim:weyl-special}
\end{claim}
\begin{IEEEproof}
Recall that the spectral radius of a matrix is the maximum absolute
value of the eigenvalues of the matrix.
For symmetric matrices, the eigenvalues are all real, and 
since \(P\) and \(Q\) are nonnegative, the Perron-Frobenius theorem ensures
that there is a positive eigenvalue which has the maximum absolute value.
Thus we have
\begin{align*}
\rho(P+Q) &= \max_{\lVert \mathbf{x} \rVert=1}{\mathbf{x}^\top (P+Q) \mathbf{x}}
= \max_{\lVert \mathbf{x} \rVert=1}\left(\mathbf{x}^\top P \mathbf{x} + \mathbf{x}^\top Q \mathbf{x}\right).
\end{align*}
Let \(\tilde{\mathbf{x}}\) be the unit vector \(\mathbf{x}\) which maximizes
\(\mathbf{x}^\top P \mathbf{x}\), i.e.,
\(\rho(P)=\max_{\lVert \mathbf{x} \rVert=1}\mathbf{x}^\top P \mathbf{x}
=\tilde{\mathbf{x}}^\top P \tilde{\mathbf{x}}\).
This gives
\begin{align*}
\max_{\lVert \mathbf{x} \rVert=1}\left(\mathbf{x}^\top P \mathbf{x} + \mathbf{x}^\top Q \mathbf{x}\right)
&\ge \tilde{\mathbf{x}}^\top P \tilde{\mathbf{x}} + \tilde{\mathbf{x}}^\top Q \tilde{\mathbf{x}} \\
&\ge \rho(P) + \min_i Q_{ii},
\end{align*}
where the second inequality follows since \(\min_iQ_{ii}\) is
the least value of \(\mathbf{x}^\top Q \mathbf{x}\) subject to \(\lVert \mathbf{x} \rVert=1\) since \(Q\) is a diagonal matrix.
This proves the lower bound of Claim~\ref{claim:weyl-special}.

For the upper bound, we have
\begin{align*}
\max_{\lVert \mathbf{x} \rVert=1}\left(\mathbf{x}^\top P \mathbf{x} +
\mathbf{x}^\top Q \mathbf{x}\right)
&\le
\max_{\lVert \mathbf{x} \rVert=1} \mathbf{x}^\top P \mathbf{x}
+\max_{\lVert \mathbf{x} \rVert=1} \mathbf{x}^\top Q \mathbf{x} \\
&\le \rho(P) + \max_i Q_{ii},
\end{align*}
which concludes the proof.
\end{IEEEproof}

\section{Proof of Claim~\ref{claim:epsilon-bound}}
\label{appendix:epsilon-bound}

Since \(\lim_{n\to\infty}\gamma(n)=0\), for any \(\epsilon > 0\), we 
can find an \(m_\epsilon\) such that for all \(n > m_\epsilon\),
\(\gamma(n) < \epsilon\).
Let \(T_{i,j}\) denote the time it takes to go from \(i\) infections to
\(j\) infections (for the first time).
Then we have
\begin{align*}
\EX[T_n] = \EX[T_{n,m_\epsilon}] + \EX[T_{m_\epsilon}].
\end{align*}
But the birth rate of the Markov chain between \(n\) and \(m_\epsilon\)
is less than \(\epsilon\) (from the definition of \(m_\epsilon\)).
So \(\EX[T_{n,m_\epsilon}]\) should be less than the expected time to go
from \(n\) to \(0\) in a Markov chain where 
all the birth rates are \(\epsilon\).
This gives us (using Claim~\ref{claim:tn-constant}):
\begin{align*}
\EX[T_n] \le \frac{\ln n}{\delta - \epsilon} + \EX[T_{m_\epsilon}] + \frac{1}{\delta-\epsilon}.
\end{align*}
Since \(\EX[T_{m_\epsilon}]\) depends only on \(\epsilon\) given a \(\gamma(\cdot)\), this concludes the proof for the second inequality.

The first inequality is relatively straightforward since 
\(\frac{\ln (n+1)}{\delta}\) is the lower bound in 
Claim~\ref{claim:tn-constant} if the birth rate was \(0\) throughout.

\section{Extinction time exponential in equilibrium point}
\label{appendix:exponential-equal}

For simplicity, we just consider the upper- and lower-bound Markov chains
using the rates from \eqref{eq:rates} defined using the \(\gamma(\cdot)\)
function here.
We expect similar arguments to hold for the network-wide epidemic as well.
Let \(\gamma(n)=\delta+\epsilon\) for all \(n \le N\)
and \(\gamma(n)=0\) for all \(n > N\).
Since this satisfies the condition of Theorem~\ref{thm:log-hitting},
we are guaranteed that the mean epidemic extinction time is logarithmic
in the initial infection size.
However, the mean extinction time also turns out to be exponential in 
\(N\), the ``equilibrium point,'' or the size of the epidemic 
where the rate of infectiousness \(\gamma(\cdot)\) goes below 
the curing rate \(\delta\).

To see this, substitute these values into the expression for
\(\EX[T_1]\) from Claim~\ref{claim:t1-gamma}.
We get
{
\begin{align*}
\EX[T_1] &=
\frac{1}{\delta}\sum_{i=1}^{N+1}\frac{1}{i}\left(\frac{\delta+\epsilon}{\delta}\right)^{i-1} \\
&\ge\frac{1}{(N+1)\delta}\sum_{i=1}^{N+1}\left(1+\frac{\epsilon}{\delta}\right)^{i-1}
\end{align*}
}
\vspace{-0.3em}
\begin{align*}
=\frac{\left(1+\frac{\epsilon}{\delta}\right)^{N+1}-1}{(N+1)\epsilon}.\quad\ 
\end{align*}
If \(N\) or \(\epsilon\) are large enough, \(\EX[T_1]\) is greater than an
exponential of the form \(a^N\) for some \(a>1\).
This implies that the mean die-out time is exponential in the 
equilibrium point \(N\).

\end{document}